\documentclass[aps,pre,twocolumn,showpacs]{revtex4}
\usepackage{amsfonts}
\usepackage{amsmath}
\usepackage{amsthm}
\usepackage{graphics}
\usepackage{graphicx}
\usepackage{subfigure}
\usepackage{amssymb}
\usepackage{graphics}
\usepackage{graphicx}
\usepackage{color}
\usepackage{subfigure}

\def\kmax{k_{\rm max}}
\newcommand\bet{{g}}
 
\newcommand\alps{{\frac{\hbar^2}{2m}}}
\newcommand\mub{{\mu}}

\newcommand\dertt[1]{ \frac{\partial{ #1}}{\partial t} }
\newcommand\gd{\mbox{${\bf \nabla}^{2}$}}
\newcommand\grad{\mbox{${\bf \nabla}$}}
\newcommand\psib{\overline{\psi}}
\newcommand{\pder}[2]{\frac{\partial{#1}}{\partial{#2}}}
\newcommand{\dert}[1]{\frac{{\rm d}{#1}}{{\rm dt}}}
\newcommand{\fder}[2]{\frac{\delta {#1}}{\delta {#2}}}

\begin{document}

\title{
Energy cascade with small-scales thermalization, counterflow metastability and anomalous velocity of vortex rings in Fourier-truncated Gross-Pitaevskii equation}
\author{Giorgio Krstulovic}
\affiliation{Laboratoire de Physique Statistique de l'Ecole Normale 
Sup{\'e}rieure, \\
associ{\'e} au CNRS et aux Universit{\'e}s Paris VI et VII,
24 Rue Lhomond, 75231 Paris, France}
\author{Marc Brachet}
\affiliation{Laboratoire de Physique Statistique de l'Ecole Normale 
Sup{\'e}rieure, \\
associ{\'e} au CNRS et aux Universit{\'e}s Paris VI et VII,
24 Rue Lhomond, 75231 Paris, France}
\date{\today}

\begin{abstract}
The statistical equilibria of the (conservative) dynamics of the Gross-Pitaevskii Equation (GPE)
with a finite range of spatial Fourier modes are characterized using a new algorithm, based on a
stochastically forced Ginzburg-Landau equation (SGLE), that directly generates grand canonical distributions. 
The SGLE--generated distributions are validated against finite-temperature GPE--thermalized states and exact (low-temperature)
results obtained by steepest descent on the (grand canonical) partition function. 
A standard finite-temperature second-order $\lambda$-transition is exhibited.

A new mechanism of GPE thermalization through a direct cascade of energy is found using initial
conditions with mass and energy distributed at large scales.
A long transient with partial thermalization at small-scales is observed before the system reaches equilibrium.
Vortices are shown to disappear as a prelude to final thermalization and their annihilation is related to the contraction 
of vortex rings due to mutual friction.
Increasing the amount of dispersion at truncation wavenumber is shown to slowdown thermalization and vortex annihilation.
A bottleneck that produces spontaneous effective self truncation with partial thermalization is characterized in the limit of large dispersive effects.

Metastable counter-flow states, with non-zero values of momentum, are generated using the SGLE algorithm. 
Spontaneous nucleation of vortex ring is observed and the corresponding Arrhenius law is characterized.
Dynamical counter-flow effects on vortex evolution are investigated using two exact solutions of the GPE:
traveling vortex rings and a motionless crystal-like lattice of vortex lines.
Longitudinal effects are produced and measured on the crystal lattice.
A dilatation of vortex rings is obtained for counter-flows larger than their translational velocity.
The vortex ring translational velocity has a dependence on temperature that is an order of magnitude
above that of the crystal lattice, an effect that is related to the presence of finite-amplitude Kelvin waves. 
This anomalous vortex ring velocity is quantitatively reproduced by assuming equipartition of energy of the Kelvin waves.
Orders of magnitude are given for the predicted effects in weakly interacting Bose-Einstein condensates  and superfluid $^4{\rm He}$.
\end{abstract}

\maketitle
\bigskip


\section{Introduction}\label{Sec:Intro}
Finite temperature superfluids are typically described as a mixture of two interpenetrating fluids \cite{Landau6Course}. 
At low temperatures the normal fluid can be neglected and Landau's two-fluids model reduces to the Euler equation
for an ideal fluid that is irrotational except on (singular) vortex lines around which the circulation of the velocity is quantized.
At finite temperature, when both normal fluid and superfluid vortices are present
(e.g. in the counterflow produced by a heat current)
their interaction, called ``mutual friction'', must also be
accounted for \cite{Vinenxxx}.

In the low-temperature regime the Gross-Pitaevskii equation (GPE) (also called the Nonlinear Schr\H{o}dinger Equation) 
is an alternative description of superfluids and Bose-Einstein Condensates (BEC) \cite{Proukakis:2008p1821}. 
The GPE is a partial differential equation (PDE) for a complex wave field that is related to the superflow's density and velocity by Madelung's 
transformation \cite{Donne}.
The (non singular) nodal lines of the complex wave field correspond to the quantum vortices that appear naturally
in this model with the correct amount of velocity circulation.
Just as the incompressible Euler equation, the GPE dynamics is known to produce \cite{Nore:1997p1333,Nore:1997p1331,Kobayashi:2005p4037,Yepez:2009p5107} an energy cascade that leads
to a Kolmogorov regime with an energy spectrum scaling as $E(k) \sim k^{-5/3}$.
This Kolmogorov regime was also experimentally observed in low temperature helium
\cite{Abid:1998p3893,Maurer:1998p4365}.
In this experimental context, let us remark that so much progress has been made that it is now possible to visualize superfluid
vortices both in the low-temperature regime and in the presence of counter flow by following the trajectories of solid
hydrogen tracers in helium \cite{Bewley:xxx,Paoletti:1545}.

Several different theories of finite-temperature effects in BEC have been proposed and, at the moment, there is no consensus on the best model \cite{Proukakis:2008p1821}. In one approach
it has been suggested that, beyond its good description of the low-temperature regime, the GPE should also be able to describe the classical equilibrium aspects of a finite-temperature
homogeneous system of ultracold gases, provided that that a projection (or truncation) on a finite number of Fourier modes
is performed \cite{Davis:2001p1475,Proukakis:2008p1821}.
Another approach to finite temperatures is the Zaremba-Nikuni-Griffin (ZNG) theory \cite{ZNG99} which couples the GPE with a Boltzmann-like equation for the thermal cloud of non-condensed particles. The ZNG theory is known to well describe the observed finite temperature decay of solitons \cite{JPB07}. It also predicts vortex motion in agreement with the standard phenomenology \cite{JPBZ09}. In the truncated GPE model the small-scales modes are in thermal equilibrium.  They play the role of the Boltzmann sector of the ZNG, somewhat like the (fast) thermalized degrees of freedom do in a standard molecular dynamics simulation. The present paper is devoted to the truncated GPE approach.

Classical truncated systems, that are similar to the truncated GPE, have a long history in the context of fluid mechanics. 
Indeed, if the (conservative) Euler equation is spectrally truncated, by keeping only a finite number of spatial Fourier
harmonics, it is well known that it admits absolute equilibrium solutions with Gaussian
statistics and equipartition of kinetic energy among all Fourier
modes \citep{LEE:1952p4100,KRAICHNAN:1955p3039,KRAICHNARH:1973p2909,OrszagHouches}.

Recently, a series of papers focused on the dynamics of convergence of the truncated Euler equation toward the absolute
equilibrium. It was found that (long-lasting) transient are obtained that are able to mimic (irreversible) viscous effects
because of the presence of a ``gas'' of partially-thermalized high-wavenumber Fourier modes 
that generates (pseudo) dissipative effects
\cite{Cichowlas:2005p1852,Bos:2006p62,Krstulovic:2008p428,Krstulovic:2009p1876,Frisch:2008p1877,Krstulovic:2009IJBC}.

The main goal of the present paper is to obtain and study finite temperature dissipative and counter flow effects by
extending to the Fourier-truncated GPE the dynamical results that were obtained in the framework of the truncated
Euler equation. We now give a short review of what is already known about the truncated GPE dynamics.

The Fourier truncated Gross-Pitaevskii equation was first introduced in the context of Bose condensation by
Davis et al. \cite{Davis:2001p1475} as a description of the classical modes of a finite-temperature
partially-condensed homogeneous Bose gas.
They considered random initial data defined in Fourier space by modes with constant modulus and random phases up to some maximum wavenumber (determined by the energy).
They found that, the numerical evolution of the truncated Gross-Pitaevskii
equation reached (microcanonical) equilibrium and that a condensation transition of the equilibrium was
obtained when the initial energy was varied.

The same condensation transition was later studied by Connaughton et al.
\cite{Connaughton:2005p1744} and interpreted as a condensation of classical nonlinear waves.
Using a modified wave turbulence theory with ultraviolet cutoff, they argued that the transition to condensation
should be subcritical. They found their theory in quantitative agreement with numerical integration
of the GPE, using the same stochastic initial conditions than those of reference \cite{Davis:2001p1475}.
However, the authors later argued that, as weak turbulence theory is expected to breakdown nearby the
transition to condensation, the subcritical nature of the transition predicted by their theory was not physical \cite{During:2009PhysicaD}.

Berloff and Svistunov \cite{Berloff:2002p3383}, starting from periodic initial conditions similar to those of 
Davis et al. \cite{Davis:2001p1475}, used a finite-difference scheme (exactly conserving energy and particle
number) to characterized the dynamical scenario of the relaxation toward equilibrium.
Using the same finite-difference scheme, Berloff and Youd \cite{Berloff:2007p423} then studied the dissipative dynamics of superfluid 
vortices at nonzero temperatures and observed a contraction of the vortex rings that followed a universal decay law.

Our main results are the followings. 
The classical absolute equilibrium of ideal fluids when generalized to GPE superfluids describes a
standard \cite{ZinnJustin:2007p3931,JAmit:1978p4003} second-order phase transition.
Long transient with energy cascade and partial small-scales thermalization are present in the
relaxation dynamics.
Dynamical counter-flow effects on vortex evolution are naturally present in the system and the vortex ring
have anomalous velocities caused by thermally  excited Kelvin waves.

The paper is organized as follows: Section \ref{Sec:Theo} is devoted to the basic theoretical background that is
needed to account for the dynamics and thermalization of the Fourier truncated GPE. 

In Sec. \ref{Sec:CharThermoEq}, the thermodynamic equilibrium is explored. The microcanonical and grand canonical
distributions are numerically shown to be equivalent.
Exact analytical expressions for the low-temperature
thermodynamic functions are obtained. A standard second-order $\lambda$ phase transition is exhibited at finite-temperature using the SGLE-generated grand canonical states.

In Sec. \ref{Sec:Cascade}, the direct energy cascade is considered as a new mechanism for GPE thermalization. 
Using initial data with mass and energy distributed at large scales, a long transient with partial thermalization at small-scales
is characterized. Vortex annihilation is observed to take place and is related to mutual friction effects. A bottleneck
producing spontaneous self truncation with partial thermalization and a time-evolving effective truncation wavenumber is
characterized in the limit of large dispersive effects at the maximum wavenumber of the simulation.

In Sec. \ref{Sec:Meta}, the new SGLE algorithm is used to generate counter-flow states, with non-zero values of momentum,
that are shown to be metastable under SGLE evolution. The spontaneous nucleation of vortex ring and the corresponding
Arrhenius law are characterized. Dynamical counter-flow effects are investigated using vortex rings and straight vortex lines arranged in crystal-like patterns. An anomalous translational velocity of vortex
ring is exhibited and is quantitatively related to the effect of thermally excited finite-amplitude Kelvin waves. 
Orders of magnitude are estimated for the corresponding effects in weakly interacting Bose-Einstein condensates and superfluid $^4{\rm He}$.

Section  \ref{Sec:Conclusion} is our conclusion. The numerical methods and low-temperature thermodynamic functions are described in an appendix.

\section{Theoretical background\label{Sec:Theo} }
This section deals with basic facts needed to understand the dynamics and thermalization of the Fourier truncated GPE. 
We first recall in section \ref{SubsubSec:GPE} the (untruncated) GPE dynamics,
its associated conserved quantities and the corresponding spectra; this material can be skipped by the reader already familiar with the GPE model of superflow \cite{Donne,Nore:1997p1331}.  The Fourier truncated GPE, its thermodynamical limit and the different statistical ensembles are then defined. 

The thermodynamics of the truncated
system is introduced in section \ref{SubSec:Thermo} using the microcanonical distribution. The canonical and grand canonical distributions are also
used as they allow to directly label the equilibrium states by temperature and particle numbers. 

A stochastically forced
Ginzburg-Landau equation (SGLE) is considered in section \ref{SubSec:SGLE} and shown to define a new algorithm that directly generates the grand canonical distributions.

\subsection{Galerkin truncated Gross-Pitaevskii equation}

\subsubsection{Conservation laws and Galilean invariance of the GPE\label{SubsubSec:GPE}}

Superfluids and Bose-Einstein condensates \cite{Abid:2003p3895,Proukakis:2008p1821} can be described at low temperature by the Gross-Pitaevskii equation (GPE) that is a partial differential equation (PDE) for the complex field $\psi$ that reads
\begin{equation} 
i\hbar\dertt{\psi}  =- \alps \gd \psi + \bet|\psi|^2\psi,  \label{Eq:GPE}
\end{equation}
where $|\psi|^2$ is the number of particles per unit volume, $m$ is the mass of the condensed particles and $g=\frac{4 \pi  \tilde{a} \hbar^2 }{m}$, with $\tilde{a}$ the $s$-wave scattering length. This equation conserves the Hamiltonian $H$, the total number of particles $N$ and the momentum ${\bf P}$ defined in volume $V$ by
\begin{eqnarray}
H&=&\int_{V} d^3 x \left( \alps |\grad \psi |^2
 +\frac{g}{2}|\psi|^4 \right)\label{Eq:defH}\\
 N&=&\int_V  |\psi|^2 \,d^3x\label{Eq:defN}\\
 {\bf P}&=&\int_V  \frac{i\hbar}{2}\left( \psi {\bf \nabla}\psib - \psib {\bf \nabla}\psi\right)\,d^3x\label{Eq:defP}.
\end{eqnarray}
It will be useful for the next sections to explicitly write the conservation law of the momentum $\partial_t  \frac{i\hbar}{2}( \psi \partial_j \psib - \psib \partial_j \psi)+ \partial_k \Pi_{kj}= 0$, where the momentum flux tensor $\Pi_{kj}$ is defined, following ref.\cite{Nore:1997p1331}, as 
\begin{equation}
\Pi_{kj}=\alps ( \partial_k \psib \partial_j \psi +\partial_k \psi \partial_j \psib) +\delta_{kj}( \frac{g}{2}|\psi|^4 -\frac{\hbar^2}{4m} \nabla^2|\psi|^2 ).\label{Eq:defMomFluxTensor}
\end{equation}

It is well known that the GPE (\ref{Eq:GPE}) can be mapped into hydrodynamics equations of motion for a compressible irrotational fluids using the Madelung transformation defined by
\begin{equation}
\psi({\bf x},t)=\sqrt{\frac{\rho({\bf x},t)}{m}}\exp{[i \frac{m}{\hbar}\phi({\bf x},t)]},\label{Eq:defMadelung}
\end{equation}
where $\rho({\bf x},t)$ is the fluid density and $\phi({\bf x},t)$ is the velocity potential such that ${\bf v}={\bf \nabla} \phi$. The Madelung transformation (\ref{Eq:defMadelung}) is singular on the zeros of $\psi$. As two conditions are required (both real and imaginary part of $\psi$ must vanish) these singularities generally take place on points in two-dimension and on curves in three-dimensions. The Onsager-Feynman quantum of velocity circulation around vortex lines $\psi=0$ is given by $h/m$.

When Eq.\eqref{Eq:GPE} is linearized around a constant  $\psi=A_0$, the sound velocity is given by $c=\sqrt{g|A_{0}|^2/m}$ with dispersive effects taking place for length scales smaller than the coherence length defined by
\begin{equation}
\xi=\sqrt{\hbar^2/2m|A_{\bf 0}|^2g }.\label{Eq:defxi}
\end{equation}
$\xi$ is also the length scale of the vortex core \cite{Nore:1997p1331,Proukakis:2008p1821}.

Following reference \cite{Nore:1997p1333} we define the total energy per unit volume $e_{\rm tot}=(H-\mu N)/V-\mu^2/2g$ where $\mu$ is the chemical potential (see section \ref{SubSec:Thermo}). Using the hydrodynamical variables, $e_{\rm tot}$ can we written as the sum of three parts:
the kinetic energy
$ e_{kin}$, the internal energy 
$e_{int}$
and the quantum energy $e_{q}$ defined by
\begin{eqnarray}
e_{\rm kin}&=&\frac{1}{V}\int d^3x \frac{1}{2}(\sqrt \rho {\bf v})^2\label{Eq:defEkin} \\
e_{\rm int}&=&\frac{1}{V}\int d^3x  \frac{g}{2m^2}\left(\rho-\frac{\mu m}{g}\right)^2\label{Eq:defEint}\\
e_{\rm q}&=&\frac{1}{V}\int d^3x \frac{\hbar^2}{2m^2}\left(\grad\sqrt{\rho}\right)^2\label{Eq:defEq}.
\end{eqnarray}
Using Parseval's theorem, one can define corresponding energy spectra:
e.g. the kinetic energy spectrum
$e_{\rm kin}(k)$ is defined as the sum over the angles %
\begin{equation}
e_{\rm kin}(k)=\int \left|\frac{1}{V} \int d^3 r e^{i {\bf r}\cdot{\bf  k}}\sqrt \rho {\bf v} \right|^2 k^2d\Omega_k,
\end{equation} 
where $d\Omega_k$ is the solid angle element on the sphere.
The energy ${e_{kin}}$ can be further decomposed into a compressible part 
$e_{\rm kin}^{\rm c}$ and an incompressible part $e_{\rm kin}^{\rm i}$ by making use of the relation $\sqrt \rho {\bf v}=(\sqrt {\rho} {\bf v})^{\rm c}+(\sqrt{\rho }{\bf v})^{\rm i}$ with ${\bf \nabla}\cdot(\sqrt{\rho }{\bf v})^{\rm i}=0$ (see \cite{Nore:1997p1331} for details).

Finally note that the GPE \eqref{Eq:GPE} is invariant under the  Galilean transformation 
\begin{equation}
\psi'({\bf x},t)=\psi({\bf x}-{\bf v}_{\rm G}t,t)\exp{\left\{  \frac{i m}{\hbar} \left[{\bf v}_{\rm G}\cdot {\bf x}- \frac{1}{2}  v_{\rm G} ^2t\right]\right\}}.\label{Eq:GaliTransf}\\
\end{equation}
Under this transformation Eqs.(\ref{Eq:defH}-\ref{Eq:defP}) transform as
\begin{eqnarray}
H' &=&\frac{1}{2}m N v_{\rm G} ^2+{\bf P}\cdot{\bf v}_{\rm G}+  H\\
N'&=&N\\
{\bf P'}&=&m N{\bf  v}_{\rm G}+ {\bf P}.\label{Eq:Ptrans}
\end{eqnarray}

\subsubsection{Definition of the Fourier truncated GPE}

For a periodical $3$D system of volume $V$ the Fourier truncated GPE is defined by performing a Galerkin truncation that consists in keeping only the Fourier modes with wavenumbers smaller than a UV cut-off $\kmax$. 

Expressing  $\psi$ in terms of the Fourier modes $A_{\bf k}$ as
\begin{equation}
\psi({\bf x},t)=\sum_k A_{\bf k}(t) e^{i {\bf k} \cdot {\bf x}}\,,{\,\rm with}\hspace{.5cm}\frac{{\bf k}}{k_{\rm min}}\in \mathbb{Z}^3 \label{eq:defFour},
\end{equation}
and where $k_{\rm min}=2\pi/V^{1/3}$ is the smallest wavenumber. The Galerkin (Fourier) truncated Gross-Pitaevskii equation (TGPE) is defined as 
\begin{equation}
-i\hbar\dertt{A_{\bf k}}  =-\frac{\hbar^2k^2}{2m} A_{\bf k}- \sum_{{\bf k_1},{\bf k_2}}A_{\bf k_1}A^*_{\bf k_2+k_1}A_{\bf k+ k_2}\label{Eq:TGPE},
\end{equation}
where the Fourier modes satisfy $A_{\bf k}=0$ if $k\ge k_{\rm max}$ and the sum is performed over all wavenumbers satisfying $|{\bf k_1}|,|{\bf k_2}|,|{\bf k_2+k_1}|,|{\bf k+k_2}|<k_{\rm max}$.
This time-reversible finite system of ordinary differential equations with a large number of degree of freedom $\mathcal{N}\sim( k_{\rm max}/k_{\rm min})^3$ also conserves the energy, number of particles and momentum.

The direct numerical evolution of the convolution in Eq.(\ref{Eq:TGPE}) would be very expensive in computational time $O(N^6)$, where $N$ is the resolution. This difficulty is avoided by using pseudo-spectral methods \cite{Got-Ors} and the non-linear term is calculated in physical space, using FFTs that reduce the CPU time to  $O(N^3\log{N})$. Introducing the Galerkin projector $\mathcal{P}_{\rm G}$ that reads in Fourier space $\mathcal{P}_{\rm G} [ A_{\bf k}]=\theta(\kmax-k)A_{\bf k}$ with $\theta(\cdot)$ the Heavside function, the TGPE (\ref{Eq:TGPE}) can be written as
\begin{equation}
i\hbar\dertt{\psi}  =\mathcal{P}_{\rm G} [- \alps \gd \psi + \bet\mathcal{P}_{\rm G} [|\psi|^2]\psi ].
\label{Eq:TGPEphys}
\end{equation}
Equation (\ref{Eq:TGPEphys}) exactly conserves energy and mass and, if it is correctly de-aliased using the $2/3$-rule \cite{Got-Ors} (dealiasing at $\kmax=\frac{2}{3}\frac{N}{2}$), it also conserves momentum (see Appendix \ref{Ap:Des&Cons} for a explicit demonstration). The Galerkin truncation also preserves the Hamiltonian structure with the truncated Hamiltonian given by 
$H=\int d^3 x \left( \alps |\grad \psi |^2 +\frac{g}{2}[\mathcal{P}_{\rm G}|\psi|^2]^2 \right).\label{Eq:HGalerkin}$

Let us remark that perhaps a more standard definition of dealiasing in Eq.\eqref{Eq:TGPEphys} could have been $\mathcal{P}_{\rm G} [|\psi|^2\psi] $ using $1/2$-rule  (dealiasing at $\kmax=\frac{1}{2}\frac{N}{2}$) rather than $\mathcal{P}_{\rm G} [\mathcal{P}_{\rm G} [|\psi|^2]\psi ]$ with the $2/3$-rule. Using the former definition removes the restriction $|{\bf k_2}|<k_{\rm max}$ on the convolution in Eq.(\ref{Eq:TGPE}). Both methods are equivalent in the partial differential equation (PDE) limit (exponential decay of energy spectrum for $k\ll\kmax$) and admit the same invariants. However the scheme of Eq.\eqref{Eq:TGPEphys} is preferable because $k_{\rm max}$ is larger at the same resolution. If dealiasing is not preformed in equation \eqref{Eq:TGPEphys} the errors in the conservation of momentum can rise up to $50\%$ in a few units of time (see Appendix \ref{Ap:Des&Cons}). In a finite difference scheme the conservation of momentum should also be checked carefully as it is bound to produce spurious effects.

Another effect caused by periodic boundary condition is that the velocity ${\bf v}_{\rm G}$ in the Galilean transformation \eqref{Eq:GaliTransf} is quantized by the relation
\begin{equation}
{\bf v}_{\rm G}=\frac{\hbar}{m}\frac{2\pi}{V^{1/3}} {\bf n_{G}}\label{Eq:discValuesVs},
\end{equation} 
where $ {\bf n_{G}}\in \mathbb{Z}^3$ and ${\bf v}_{\rm s}$ becomes continuous only in the limit $\hbar/(mV^{1/3})\to0$. The Galilean invariance is slightly broken by the TGPE \eqref{Eq:TGPE} because of modes close to the truncation wavenumber $\kmax$. However it is recovered in the PDE limit where high wavenumber modes are converging exponentially and also in the thermodynamic limit: $\frac{k_{\rm max}} {k_{\rm min}}\to \infty$ defined below because the offending terms represent only a surface effect in Fourier space. 

\subsubsection{Thermodynamical limit and statistical ensembles}

Let us first notice that the energy $H$, the number of particles $N$ and the momentum ${\bf P}$ in Eqs.(\ref{Eq:defH}-\ref{Eq:defP}) are all proportional to the total number of modes $\mathcal{N}\sim\kmax^3 V$ and therefore are all extensive quantities. Also note that by definition of the coherence length \eqref{Eq:defxi}, the number $\xi\kmax$ determines the amount of dispersion at truncation wavenumber in the system. 

The thermodynamic limit $V\to\infty$ of the truncated Gross-Pitaevskii system is thus defined as the limit 
\begin{equation}
\mathcal{N}\to\infty\hspace{2mm}, \hspace{4mm}\xi\kmax= {\rm constant},\label{Eq:ThermoLimit}
\end{equation}
in order to obtain equivalent systems. In this limit the relevant thermodynamic variables are the intensive quantities $H/V$, $N/V$ and ${\bf P}/V$. In practice, to perform numerical computations we will fix the volume to $V=(2\pi)^3$ and we will vary $\kmax$ (see paragraphs before section \ref{Sec:CharThermoEq}).

Let us define, as usual the microcanonical ensemble \cite{Landau5Course} by the probability $dw$ of finding the system in states with given values of energy $H_{\rm in}$, number of particles $ N_{\rm in}$ (the subscript ``in'' stands for initial data) and momentum ${\bf  P_{\rm in}}$ given by:
\begin{equation}
dw={\rm constant}\,e^S\delta(H-H_{\rm in})\delta(N-N_{\rm in})\delta^3({\bf P}-{\bf P_{\rm in}})dHdNd^3P\label{Eq:probMicro},
\end{equation}
where $S=\log{\Gamma}$ is the entropy with $\Gamma$ the number of accessible micro-states. 

Microcanonical statistical states can be obtained numerically by time-integrating the TGPE until the system reaches thermodynamic equilibrium \cite{Davis:2001p1475,Connaughton:2005p1744}. These thermalized states are formally determined by the control values $H_{\rm in}$, $ N_{\rm in}$ and ${\bf P_{\rm in}}$ that are set in the initial condition. It has been shown in references \cite{Davis:2001p1475,Connaughton:2005p1744} by varying the values of $H_{\rm in}$ that TGPE present a phase transition analogous to the one of Bose-Einstein condensation, where the amplitude at $0$-wave-number $A_{\bf 0}$ vanish for finite values of $H_{\rm in}$.  Let us remark that an explicit expression of $dw$ or $S$ cannot be easily obtained in the microcanonical ensemble and therefore the temperature is not easily accessible.

A simple way to explicitly control the temperature is to use the canonical or grand canonical formulation. The grand canonical distribution probability is given by a Boltzman weight 
\begin{eqnarray}
\mathbb{P}_{\rm st}&=&\frac{1}{\mathcal{Z}}e^{-\beta F}\label{Eq:StatProb}\\
F& =&H-\mu N-{\bf W}\cdot {\bf P}\label{Eq:defF},
\end{eqnarray}
where $\mathcal{Z}$ is the grand partition function, $\beta$ is the inverse temperature and $\mu$ is the chemical potential. In what follows we will refer to ${\bf W}$ as the counterflow velocity. 

Note that when ${\bf W}=0$, $F=H-\mu N$ and the statistic weight of distribution (\ref{Eq:StatProb}) corresponds to the so-called $\lambda \phi^4$ theory studied in second order phase transitions \cite{JAmit:1978p4003,ZinnJustin:2007p3931}. This point will be further discussed in subsection \ref{SubSection:Lambda-Trans}.

Finally remark that the states with ${\bf W} \ne 0$ are obtained, in the thermodynamic limit, by a Galilean transformation of the basic ${\bf W}=0$ state (see below Eq.\eqref{Eq:psimeta}). However, for finite size systems, because of the quantification of the Galilean transformation (Eqs.\eqref{Eq:GaliTransf} and  \eqref{Eq:discValuesVs}) new metastable states with counterflow appear. These metastable states and their interactions with vortices will be studied in detail below in section \ref{subsec:Meta}.

In the grand canonical ensemble (\ref{Eq:StatProb}-\ref{Eq:defF}) the mean energy $\overline{H}$, number of particles $\overline{N}$ and momentum $\overline{\bf P}$ are easily obtained  by defining the grand canonical potential 
\begin{equation}
\Omega=-\beta^{-1}\log{\mathcal{Z}}\label{Eq:grandCanoPot}
\end{equation}
and using the relations
\begin{equation}
\overline{N}=-\pder{\Omega}{\mu}\,,\hspace{3mm}\overline{\bf P}=-\pder{\Omega}{{\bf P}},\hspace{3mm}\overline{H}=\pder{\Omega}{\beta}+\mu\overline{N}+{\bf W}\cdot {\bf P}.
\end{equation}

Observe that the microcanonical states \eqref{Eq:probMicro} are characterized by the values $H_{\rm in}$, $N_{\rm in}$ and ${\bf P_{\rm in}}$. On the other hand, the grand canonical states are controlled by the conjugate variables: $\beta$, $\mu$ and ${\bf W}$. The different statistical ensembles are expected to be equivalent in the thermodynamics limit \eqref{Eq:ThermoLimit} and therefore
\begin{equation}
H_{\rm in}=\overline{H}\,,\hspace{3mm}N_{\rm in}=\overline{N}\,,\hspace{3mm}{\bf P_{\rm in}}=\overline{\bf P},
\end{equation}
in this limit. The equivalence of ensembles will be numerically tested below in subsection \ref{SubSec:CompEnsembles}.

In the grand canonical ensemble, the pressure $p$ is usually defined from the grand canonical potential \eqref{Eq:grandCanoPot} by the relation \cite{Landau5Course} $\Omega=-pV$. This definition presents two problems in the TGPE system. First, due to classical statistics $\Omega$ has a logarithmic divergence at $\beta=\infty$. Second, this definition does not coincide with the standard relation in fluid dynamics involving the diagonal part of the momentum flux tensor $ \Pi_{ij}$ (see Eq.(\ref{Eq:defMomFluxTensor})). Both these problems can be solved by considering the total number of modes as a new thermodynamics variable, as we will see in the next section.

\subsection{Thermodynamics of the truncated system \label{SubSec:Thermo}}

When a Galerkin truncation is performed on a system a new variable $k_{\rm max}$ explicitly appears. One thus find that the thermodynamic potentials depend on the total number of modes. Denoting $\lambda_\mathcal{N}$ the conjugate variable to the total number of modes $\mathcal{N}$ the standard thermodynamic relation for the energy easily generalizes as
  \begin{equation}
 dE = -p dV +T dS + \mu dN + \lambda_\mathcal{N} d\mathcal{N}+ {\bf W\cdot}d{\bf P}\label{Eq:dE}
 \end{equation}
with $S$ the entropy and where we have included the total momentum dependence $d{\bf P}$. As in Landau two-fluid model \cite{Landau6Course} Eq.\eqref{Eq:dE} is written in a system of reference where ${\bf v_{\rm s}}=\overline{\grad \phi}={\bf 0}$ (the bar standing for some ensemble average) and $E=\overline{H}$ is the macroscopic energy. \footnote{The Galilean invariant expression of ${\bf W}$ is ${\bf V_{\rm n}-V_{\rm s}}$ (see sec.\ref{Sec:Meta})}. We will omit the bar over the others microscopic quantities. Note that the Fourier modes formally play the role of ``particles'' and $ \lambda_\mathcal{N}$ is formally the ``chemical potential'' associated to those ``particles''.
 
The thermodynamic potentials can be easily generalized to take in to account the new variables. It is useful to define the Gibbs potential $G$, grand canonical $\Omega$ and a generalized grand canonical potential $\Omega'$ (with a Legendre transformation on $\mathcal{N}$) as
 \begin{eqnarray}
 G &=&E-TS+pV-{\bf W\cdot}{\bf P}\label{Eq:Gibbs}\\
 \Omega&=&E-TS-\mu N-{\bf W\cdot}{\bf P} \label{Eq:PartialOmega}\\
 \Omega'&=&E-TS-\mu N- \lambda_\mathcal{N} \mathcal{N}-{\bf W\cdot}{\bf P} \label{Eq:Omega}
 \end{eqnarray}
 from where their respective variations follows:
 \begin{eqnarray}
 dG &=&Vdp-SdT+\mu dN + \lambda_\mathcal{N} d\mathcal{N}-{\bf P}\cdot{\bf dW}\label{Eq:dG}\\
 d\Omega&=&-pdV-SdT-Nd\mu+\lambda_\mathcal{N}d\mathcal{N}-{\bf P}\cdot{\bf dW}\label{Eq:dPartialOmega}\\
d\Omega'&=&-pdV-SdT-Nd\mu-\mathcal{N}d\lambda_\mathcal{N}-{\bf P}\cdot{\bf dW}.\label{Eq:dOmega}
 \end{eqnarray}
  
Based on standard arguments of extensive variables \cite{Landau5Course} and noting that $\lambda_\mathcal{N}$ and ${\bf W}$ are intensive variables
we find the standard formula of the Gibbs potential with two types of particles
\begin{equation}\label{Eq:GibbsExt}
G=\mu N+\lambda_\mathcal{N} \mathcal{N}.
\end{equation}

Using Eqs.(\ref{Eq:Gibbs}) and (\ref{Eq:GibbsExt}) in Eqs.(\ref{Eq:PartialOmega}) and  (\ref{Eq:Omega}) we find
\begin{equation}
\Omega=-pV+\lambda_\mathcal{N} \mathcal{N}\label{Eq:OmegaExt},\hspace{1cm}\Omega'=-pV
\end{equation}

The relations (\ref{Eq:dE}-\ref{Eq:OmegaExt}) determine all the thermodynamic variables and potentials. For instance the pressure $p$ can be obtained from Eq.(\ref{Eq:dPartialOmega}), Eq.(\ref{Eq:dOmega}) or Eq.(\ref{Eq:OmegaExt}) by
\begin{equation}\label{Eq:PresThermo}
p=-\left.\frac{\partial \Omega}{\partial V}\right|_{T,\mu,\mathcal{N},{\bf W}}=-\frac{\Omega-\lambda_\mathcal{N} \mathcal{N}}{V}=-\frac{\Omega'}{V}
\end{equation}
where
 $\lambda_\mathcal{N} =\left.\frac{\partial \Omega}{\partial \mathcal{N}}\right|_{V,T,\mu,{\bf W}}$.

We proceed now to show that thermodynamic definition (\ref{Eq:PresThermo}) of the pressure coincides with the standard relation in fluid dynamics. In order to make explicit the dependence of the energy $H$ on the volume $V$ let us define the dimensionless space variables $\tilde{x}=x/V^{\frac{1}{3}}$ and $\tilde{\psi}=V^{1/2}\psi$. Expressed in term of these variables the Hamiltonian (\ref{Eq:defH}) reads $H=\int d^3 \tilde{x} \left( \alps \frac{1}{V^{\frac{2}{3}}} |\tilde{\grad} \tilde{\psi} |^2
 +\frac{1}{V}\frac{g}{2}|\tilde{\psi}|^4 \right)$. Taking the derivative with respect to $V$ and reintroducing $x$ and $\psi$ yields
\begin{equation}
\frac{\partial H}{\partial V}=-\frac{1}{V}\int d^3 x \left( \alps \frac{2}{3} |\grad \psi |^2
 +\frac{g}{2}|\psi|^4 \right).\label{Eq:dHsdV}
\end{equation}
This expression corresponds to the spatial average of the the diagonal part of $\Pi_{ik}$ (see Eq.(\ref{Eq:defMomFluxTensor})).
As by definition $E=\overline{H}$ and the derivative has been implicitly done at constant total number of modes  and momentum we find, using the thermodynamic relation (\ref{Eq:dE}) and Eq.(\ref{Eq:dHsdV}), that  the pressure satisfies
\begin{equation}
p=-\left.\frac{\partial E}{\partial V}\right|_{S,N,\mathcal{N},{\bf P}}=-\left.\overline{\frac{\partial H}{\partial V}}\right|_{N,\mathcal{N},{\bf P}},
\end{equation}
 where the second equality holds for adiabatic compressions \cite{Landau5Course}.

Finally by replacing $\Omega$ in Eq.(\ref{Eq:PartialOmega}) we obtain the thermodynamic relation
\begin{equation}
E+pV-\mu N -{\bf W\cdot}{\bf P}=T S+ \lambda_\mathcal{N}\mathcal{N}\label{Eq:ThermoRelation}.
\end{equation}
Let us remark that, in a classical system, the entropy is defined up to an additive constant related to the normalization of the phase-space.
However the quantity $T S+ \lambda_\mathcal{N}\mathcal{N}$ is completely determined because each term in the left hand side of Eq.(\ref{Eq:ThermoRelation}) is well defined. By the same arguments $d\left(\mathcal{N}\lambda_\mathcal{N}/T\right)$ is also a completely determined quantity. If the variable $\mathcal{N}$ had not been taken into account, the corresponding pressure would be $-\Omega/V$ and therefore wrongly defined and depending on the normalization constant. The grand canonical potential $\Omega$ will be explicitly obtained at low-temperature in subsection \ref{sec:LowTcalulations} where the above considerations can be explicitly checked.

\subsection{Generation of grand canonical distribution using a stochastic Ginzburg-Landau equation\label{SubSec:SGLE}}

Grand canonical equilibrium states are given by the statistics (\ref{Eq:StatProb}-\ref{Eq:defF}). They cannot be easily obtained because the Hamiltonian $H$ in Eq.\eqref{Eq:defH} is not quadratic and therefore the statistical distribution is not Gaussian. Nevertheless it is possible to construct a stochastic process that converges to a stationary solution with equilibrium distribution (\ref{Eq:StatProb}-\ref{Eq:defF}). This process is defined by a Langevin equation consisting of a stochastic Ginbzurg-Landau equation (SGLE) that reads
\begin{eqnarray}
\hbar\dertt{A_{\bf k}} & =&-\frac{1}{V}\pder{F}{A_{\bf k}^*}+\sqrt{\frac{2 \hbar}{V\beta }}\,\hat{\zeta}({\bf k},t)\label{Eq:SGLR}\\
\langle\zeta({\bf x},t)\zeta^*({\bf x'},t')\rangle&=&\delta(t-t') \delta({\bf x}-{\bf x'})\label{Eq:noise},
\end{eqnarray}
where $F$ is defined in Eq.(\ref{Eq:defF}) and $\hat{\zeta}({\bf k},t)$ is the ($\kmax$-truncated) Fourier transform of the gaussian white-noise $\zeta({\bf x},t)$ defined by Eq.(\ref{Eq:noise}). The Langevin equation (\ref{Eq:SGLR}-\ref{Eq:noise}) explicitly reads in physical space
\begin{eqnarray}
\nonumber\hbar\dertt{\psi} &=&\mathcal{P}_{\rm G} [ \alps \gd \psi +\mub \psi- \bet\mathcal{P}_{\rm G} [|\psi|^2]\psi-i\hbar{\bf W}\cdot{\bf\nabla} \psi ]\\
&&\hspace{1.5cm}+\sqrt{\frac{2 \hbar}{V\beta }}\mathcal{P}_{\rm G} [ \zeta({\bf x},t)] \label{Eq:SGLRphys}.
\end{eqnarray}

In the $\beta\to\infty$ limit Eq.\eqref{Eq:SGLRphys} reduces to the advective real Ginzurg-Landau equation (up to a redefinition of $\mu$) that was introduced in reference \cite{Nore:1997p1331}. This equation has the same stationary solutions of than the TGPE \eqref{Eq:TGPEphys} in a system of reference moving with velocity ${\bf W}$. When the term $\mu \psi$ is also included in the TGPE it has, because of particle number conservation,  the only effect of adding a global time-dependent phase factor to the solution.

The probability distribution $\mathbb{P}\left[\{A_{\bf k},A_{\bf k}^*\}_{\bf k<k_{\rm max}}\right]$ of the stochastic process defined by Eqs.(\ref{Eq:SGLR}-\ref{Eq:noise}) can be shown to obey the following Fokker-Planck equation \cite{Van-Kampen-Chem, LivreVertET}
\begin{equation}
\dertt{\mathbb{P}}=\sum_{\bf k<k_{\rm max}}\pder{}{A_{\bf k}}\left[\frac{1}{V\hbar}\pder{F}{A_{\bf k}^*}\mathbb{P}+\frac{1}{V\hbar\beta}\pder{\mathbb{P}}{A_{\bf k}^*}\right]+c.c\,.\label{Eq:FPNGLR}
\end{equation}
It is straightforward to demonstrate that the probability distribution \eqref{Eq:StatProb} is a stationary solution of Eq.\eqref{Eq:FPNGLR}, provided that $\beta F$ is a positive defined function of $\{A_{\bf k},A_{\bf k}^*\}_{\bf k<k_{\rm max}}$. 

If one wishes to directly control, instead of the chemical potential $\mu$, the value of the number of particles $N$ or the pressure $p$, the SGLE must we supplied with one of two \emph{ad-hoc} equation for the chemical potential. These equation simply read
\begin{eqnarray}
\dert{\mu}&=&-\nu_N(N-N^*)/V\label{Eq:muAtRho}\\
\dert{\mu}&=&-\nu_p( p-p^*)\label{Eq:muAtPres}
\end{eqnarray}
where the pressure $p$ is computed as $p=-\pder{H}{V}$ (see Eq.\eqref{Eq:dHsdV}). Equation \eqref{Eq:muAtRho} controls the number of particles and fixes its mean value to the control value $N^*$. Similarly Eq.\eqref{Eq:muAtPres} controls the pressure and fixes its value at $p^*$. Equations (\ref{Eq:muAtRho}-\ref{Eq:muAtPres}) are not compatible and they must not be used simultaneously. Depending on the type of temperature scans, the SGLE must be used together with either Eq.\eqref{Eq:muAtRho}, Eq.\eqref{Eq:muAtPres} or alone with a fixed value of $\mu$.

In the rest of this paper we will perform several numerical simulations of the TGPE \eqref{Eq:TGPE} and SGLE \eqref{Eq:SGLRphys}. For numerics, the parameters in SGLE (omitting the Galerkin projector $\mathcal{P}_{\rm G}$) will be rewritten as
\begin{eqnarray}
\nonumber\dertt{\psi} &=& \alpha_0\gd \psi +\Omega_{0} \psi- \beta_{0}|\psi|^2\psi-i{\bf W}\cdot{\bf\nabla} \psi +\sqrt{\frac{k_{B}T}{\alpha_0}}\zeta,
\end{eqnarray}
with similar changes for TGPE.

In terms of $\alpha_0$, $\Omega_{0}$ and $\beta_{0}$ the physical relevant parameters are the coherence length $\xi$ and the velocity of sound $c$ defined in section \ref{SubsubSec:GPE} (see Eq.\eqref{Eq:defxi} and the text before). They can be are expressed as
\begin{equation}
\xi=\sqrt{\alpha_0/\Omega_{0}\,}, \hspace{.5cm} c=\sqrt{2\alpha_{0}\beta_{0}\rho^*}
\end{equation}
with $\rho^*=\Omega_{0}/\beta_{0}$. The value of the density at $T=0$ set to $\rho^*=1$ in all the simulations presented below.  
In order to keep the value of intensive variables constant in the thermodynamic limit \eqref{Eq:ThermoLimit}, with $V$ constant and $\kmax\to \infty$ the inverse temperature is expressed as $\beta=1/k_{\mathcal{N}}T$ where $k_{\mathcal{N}}=V/\mathcal{N}$. With these definitions the temperature $T$ has units of energy per volume and $4\pi\alpha_0$ is the quantum of circulation.

With $\xi$ fixed, the value of $\xi/c$ only determine a time-scale. The velocity of sound is (arbitrarily) set to $c=2$ and the different runs presented below are obtained by varying only the coherence length $\xi$, the temperature $T$, the counterflow velocity ${\bf W}$ and the UV cut-off wavenumber $\kmax$. The number $\xi\kmax$ is kept constant (at the value  $\xi\kmax=1.48$) when the resolution is changed, except in section \ref{SubSection:SelfTrunc} where dispersive effects are studied (using a larger $\xi\kmax$). This choice of $\xi$ ensures that vortices are well resolved (e.g. compare Fig.\ref{Fig:TGb}.a below with Fig.12 of ref.\cite{Nore:1997p1331}). In the present work we use resolutions varying from $32^3$ to $512^3$ colocation points ($\kmax=10$ to $170$ respectively). Finally in all numerical results the energy and momentum are presented per unit of volume $V=(2\pi)^3$ and the control values of number of particles and pressure in Eqs.(\ref{Eq:muAtRho}-\ref{Eq:muAtPres}) are set to $mN^*/V=\rho^*=1$ and $p^*=c^{2}{\rho^*}^2/2=2$. Numerical integrations are performed with periodic pseudo-spectral codes and the time-stepping schemes are Runge-Kutta of order 4 for TGPE and implicit Euler for SGLE.

\section{Characterization of thermodynamic equilibrium\label{Sec:CharThermoEq}}

In this section, the thermodynamic equilibrium is explored and characterized.
The microcanonical and grand canonical
distributions are first shown to be numerically equivalent in a range of temperatures by comparing the statistics of GPE and SGLE generated states in section \ref{SubSec:CompEnsembles}.
The steepest descent method is then applied to the grand partition function in section \ref{sec:LowTcalulations} to obtain exact analytical expressions for the low-temperature thermodynamic functions. The basic numerical tools are validated by reproducing these low-temperature results.
In section \ref{SubSection:Lambda-Trans} a standard finite-temperature second-order $\lambda$ phase transition is exhibited using the SGLE-generated 
grand canonical states and the deviations to low-temperature equipartition are characterized.

\subsection{Comparison of microcanonical and grand canonical states\label{SubSec:CompEnsembles}}

We now numerically compare the statistics of the grand canonical states produced by the new algorithm SGLE to the statistics of the microcanonical states obtained by long-time integrations of TGPE. The coherence length is set to $\xi=\sqrt{2}/10$ and $32^3$ collocation points are used ($\kmax=10$). The initial condition for the TGPE runs are chosen with random phases in a similar way than in references \cite{Davis:2001p1475,Connaughton:2005p1744}. We obtain low, medium and high values of the energy with constant density $\rho=mN/V=1$ (see table \ref{Table:CompNLS-NGLR}). 
\begin{table}[htdp]
\caption{Parameters of TGPE initial condition and time steps.}
\begin{center}
\begin{tabular}{|c|c|c|c|}
\hline
$H$ & $T$  &TGPE time steps & SGLE time steps\\
\hline
\hline
$0.09$ & $0.09$ &$40000$ &  $9600$  \\
$0.5$ & $0.5$      &$20000$ &  $9600$  \\
$1.96$ & $1.8$    &$20000$ &  $9600$  \\
$4.68$ & $4$       &$20000$ &  $5000$  \\
\hline
\end{tabular}
\end{center}
\label{Table:CompNLS-NGLR}
\end{table}%

\begin{figure}[htbp]
\begin{center}
\includegraphics[height=7cm]{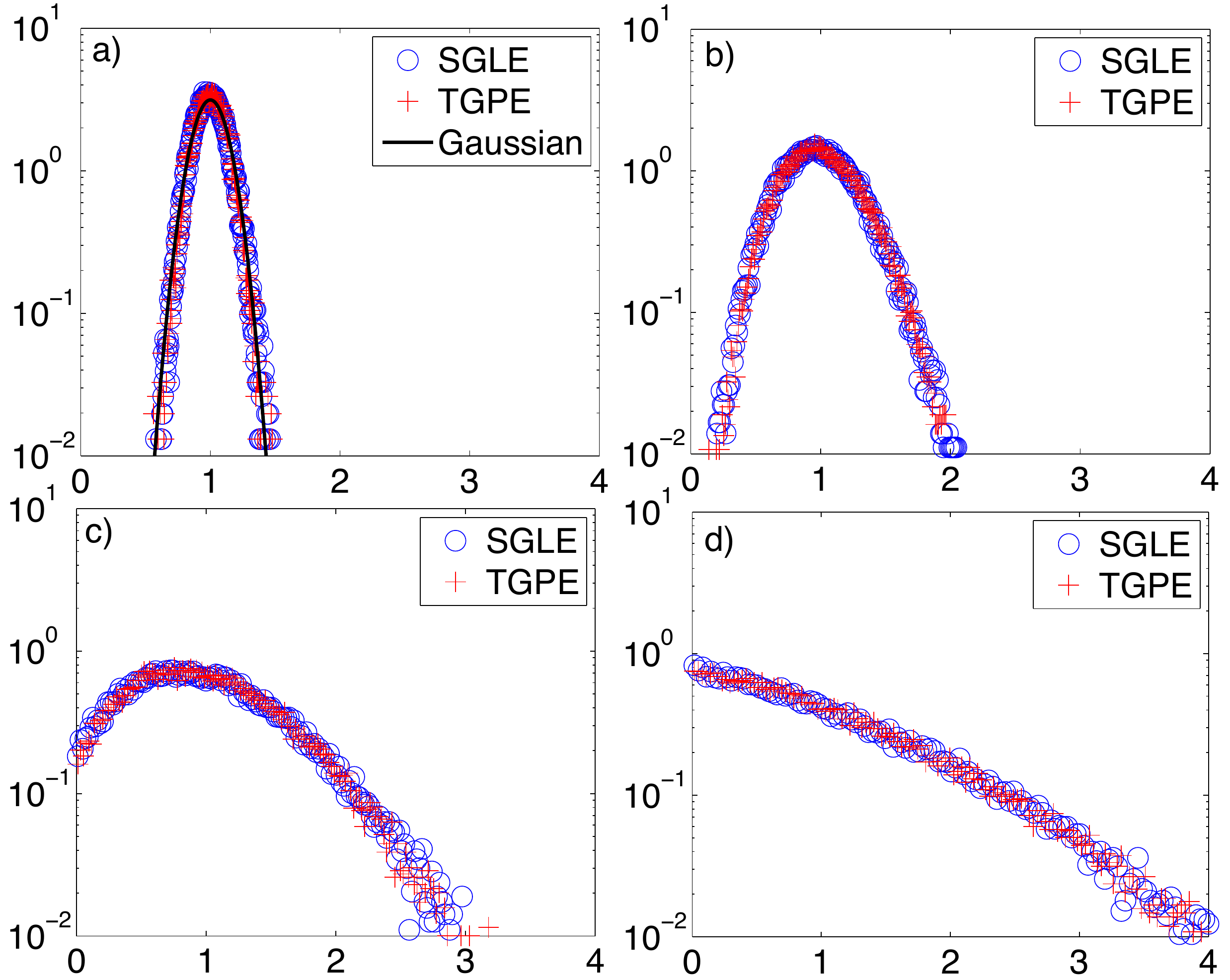}
\caption{(Color online) Comparison of density histograms obtained by SGLE and TGPE dynamics  ($\xi=2/10\sqrt{2}$ and resolution $32^3$) with energy equal to a) $H=0.09$, b) $H=0.51$, c) $H=1.96$ and d) $H=4.68$ (see table \ref{Table:CompNLS-NGLR}). The solid line in a) is a Gaussian of standard deviation $\bar{\delta\rho^2}=0.016$ (see below Eq.\eqref{Eq:deltarho}) computed with the low-temperature calculations of section \ref{sec:LowTcalulations}.}
\label{Fig:HisRho}
\end{center}
\end{figure}

To compare with the SGLE generated statistics a scan in temperature at constant density $\rho=1$ is performed in order to obtain the temperature corresponding to the energies of the TGPE runs.  Using the thermalized final states obtained from TGPE and converged final states of SGLE histograms of the of the density  $\rho(x)$ in physical space are confronted in Fig.\ref{Fig:HisRho}. They are found to be in excellent agreement.

Observe that when the energy (or temperature) increases more weight becomes apparent on the histograms near $\rho=0$, indicating the presence of vortices. The Gaussian character of the histogram in Fig.\ref{Fig:HisRho} a  motivates the low-temperature calculation of the next section. Observe that, even at this relatively low $32^3$ resolution, the thermodynamic limit has been reached in the sense that the micro and grand canonical distribution coincide. We can thus safely use, at $32^3$ and higher resolutions, the SGLE to prepare absolute equilibria of the TGPE.

Let us finally remark that the SGLE numerically converges much more rapidly toward absolute equilibrium than the TGPE, as displayed on table \ref{Table:CompNLS-NGLR}. Also taking into account the (computationally expansive) accurate conservative temporal scheme needed for the integration of the TGPE, the SGLE yields a (very) large economy of the CPU time needed to reach equilibrium. On the local machines where these computations were performed the SGLE was typically more than $10$ times faster than TGPE.

\subsection{Low-temperature calculation \label{sec:LowTcalulations}}

The gaussian histogram of Fig.\ref{Fig:HisRho}.a strongly suggest that some quadratic approximation should be able
to obtain exact analytical expressions for the thermodynamic functions at low temperature.
In this section we use a such an approximation to compute the grand partition function $\mathcal{Z}$ and the grand canonical potential \footnote{Grand canonical computations avoid difficulties that are present in the canonical ensemble with the explicit conservation of the number of particles (see section 2.2 of reference \cite{Zagrebnov:2001p3303} and references therein)}
 $\Omega=-\beta^{-1}\log{\mathcal{Z}}$ defined in \eqref{Eq:PartialOmega}. 

The first step is to express the energy $F$ of Eq.\eqref{Eq:defF} in terms of the Fourier amplitudes $A_{\bf k}$. This leads to a non quadratic function $ F\left[A_{\bf k},A_{\bf k}^*\right]$ explicitly given in appendix \ref{App:Low-T} (Eqs.\ref{energie2}-\ref{Eq:PinFourier}). The grand partition function is a product integral over all the Fourier amplitudes
\begin{equation}
\mathcal{Z}(\beta,\mu,{\bf W})=V^\mathcal{N}\int \frac{dA_{\bf 0}d{A_{\bf 0}^*}}{2\pi}\prod_ {\bf k<\kmax}\frac{dA_{\bf k}d{A^*}_{\bf k}}{2\pi}e^{-\beta F\left[A_{\bf k},A_{\bf k}^*\right]}.\label{Eq:defZformal}
\end{equation}
The integrals in (\ref{Eq:defZformal}) cannot be done explicitly, however it is possible to give a low-temperature approximation using the method of \emph{steepest descent} \cite{Mathews:1970p5033,ZinnJustin:2007p3931}. 
We also add to $F$ an \emph{external field} with value $-\mu_{0}| A_{\bf 0}|^2V$ in order to explicitly obtain the mean value of condensate Fourier mode $\bar{| A_{\bf 0}|^2}$ by direct differentiation. The physical partition function is finally obtained by setting $\mu_0=0$. The integrals are dominated by the saddle-point determined by $\pder{F}{A_{\bf k}^*}-\mu_{0}A_{\bf 0}V\delta_{\bf k,0}=0$ that yields the solution (see Eqs.\eqref{Eq:col2} and \eqref{Eq:col3})
\begin{equation}
\left.g| A_{\bf 0}|^2\right|_{\rm sp}=\mu+\mu_{0}\, \hspace{1cm}  \left.A_{\bf k}\right|_{\rm sp}=0\,\,{\rm for  }\, {\bf k}\,\neq{\bf 0}, \label{Eq:solcol}
\end{equation}
where the subscript ``sp'' stands for saddle-point. Note that in general $\bar{| A_{\bf 0}|^2}\neq \left.| A_{\bf 0}|^2\right|_{\rm sp}$ and the mean value is equal to the saddle-point one only at $T=0$. Other solutions that can be obtained when ${\bf W}\neq0$ will be discussed in detail in section \ref{Sec:Meta}.

At the saddle-point \eqref{Eq:solcol} $A_{\bf k}$ vanishes for ${\bf k}\,\neq{\bf 0}$. We thus need to keep only quadratic terms in $A_{\bf k}$ to obtain the  low-temperature approximation.
Using the notation ${\bf p}=\hbar{\bf k}$, at leading order $F$ can be rewritten as $F=F_{0}+F_{1}+F_{2}$ with
\begin{eqnarray}
F_{0}&=&V(\frac{ g}{2}|A_{\bf 0}|^4-\mu |A_{\bf 0}|^2)\\
F_{1}&=&V\sum_{\bf p\neq0}( \frac{p ^2}{2 m} -\mu+2g|A_{\bf 0}|^2-{\bf W}\cdot {\bf p})|A_{\bf p}|^2\\
F_{2}&=&V\frac{ g}{2} \sum_{\bf p\neq0}{A^*}_{\bf 0}^2A_{\bf p}A_{\bf -p}+{A}_{\bf 0}^2A^*_{\bf p}A^*_{\bf -p}.
\end{eqnarray}
In order to obtain the low-temperature partition function we need to compute the determinant of the matrix $\frac{\partial^2 F}{\partial A_{\bf p}\partial A_{\bf q} }-\mu_{0}V\delta_{\bf p,0}\delta_{\bf q,0}$. This determinant can be obtained by making use of the Bogoliubov transformation
\begin{equation}
A_{\bf p}=u_pB_{\bf p}+v_pB_{\bf -p}^*\label{Eq:transBogo}
\end{equation}
with $u_p=\frac{A_{\bf 0}}{|A_{\bf 0}|}\frac{1}{\sqrt{1-L_{p}^2}}$,  $v_p=\frac{A_{\bf 0}}{|A_{ \bf 0}|}\frac{L_{ p}}{\sqrt{1-L_{p}^2}}$ and where $L_{p}$ is determined by imposing the diagonalization of $F-\mu_{0}| A_{\bf 0}|^2V$. $L_{p}$ is explicitly given in Eq.\eqref{Eq:eqLp}.
It is easy to show that (\ref{Eq:transBogo}) is a canonical transformation and that the normalization condition of the corresponding Poisson bracket implies $|u_{ p}|^2-|v_{ p}|^2=1$.

Expressing $F$ in the Bogoliubov basis we obtain
\begin{equation}
F=V\left[\frac{ g}{2}|A_{\bf 0}|^4-\mu |A_{\bf 0}|^2+\sum_{\bf p\neq0}\left(\epsilon(p;\mu,\mu_{0})-{\bf W}\cdot {\bf p}\right) |B_{\bf p}|^2\right]\label{Eq:Fbogu}
\end{equation}
with the dispersion relation (see appendix \ref{App:Low-T})
\begin{equation}
\epsilon(p;\mu,\mu_{0})=\sqrt{\left(\mu+2\mu_{0}+\frac{p^2}{2 m} \right)^2-(\mu+\mu_{0})^2}.\label{Eq:RelDispSaddle}
  \end{equation}
Let us recall that the exited modes  $B_{\bf p}$ are called phonons in quantum mechanics. 
In the present classical case, because of classical statistics and quadratic Hamiltonian, there will be equipartition among phonon modes. Replacing the value of the chemical potential by its saddle-point expression $\mu=\left.g| A_{\bf 0}|^2\right|_{\rm sp}$  (at $\mu_0=0$), Eq.\eqref{Eq:RelDispSaddle} yields the (standard, see ref.\cite{Landau9Course}) Bogoliubov dispersion relation $\epsilon(p)=p\sqrt{\frac{g|A_{\bf 0}|^2}{m}+\frac{p^2}{4m^2}}$. Note that $\epsilon( p)$ can also be directly obtained from the GPE by expressing $\psi$ in hydrodynamics variables, using the Madelung transformation   (\ref{Eq:defMadelung}) and linearizing around an homogenous density $\rho_0=m|A_{\bf 0}|^2$  \cite{Nore:1997p1331}. 

The partition function now trivially factorizes in independent parts $\mathcal{Z}(\beta,V,\mu,{\bf W},\mathcal{N},\mu_{0})=\mathcal{Z}_{0}(\beta,\mu,\mu_{0})\prod_ {\bf p\neq0}\mathcal{Z_ {\bf p}}(\beta,\mu,{\bf W},\mu_{0})$ where
\begin{eqnarray}
\mathcal{Z}_{0}(\beta,V,\mu,\mu_{0})&=&\sqrt{2} \pi ^3 \sqrt{\frac{V}{g  \beta }} e^{\frac{V \beta ( \mu+\mu_{0}) ^2}{2g}}\label{Eq:defZ0}\\
\mathcal{Z_ {\bf p}}(\beta,\mu,{\bf W},\mu_{0})&=&\frac{1}{\beta(\epsilon(p;\mu,\mu_{0})-{\bf W}\cdot {\bf p})}\label{Eq:defZp}
\end{eqnarray}
The  total number of modes $\mathcal{N}=\sum_k 1$ and the grand canonical potential
\begin{eqnarray}
\Omega(\beta,V,\mu,{\bf W},\mathcal{N})&=&-\beta^{-1}\left[\log\mathcal{Z}_0+\sum_{\bf p\neq0}\log{\mathcal{Z_ {\bf p}}}\right]
\label{Eq:OmegDisc}
\end{eqnarray}
 are sums over all wave-numbers from which all thermodynamic quantities can be directly obtained by using the thermodynamic relation (\ref{Eq:dPartialOmega}).

Replacing the sum by an integral the expression for the number of modes reads
\begin{equation}
\mathcal{N}=\int_0^{P_{\rm max}}\frac{p^2 V}{2 \pi ^2   \hbar 
   ^3}\,dp=\frac{P_{\rm max}^3 V}{6 \pi ^2  \hbar ^3}\label{Eq:Nmodes}.
\end{equation} 
Setting ${\bf W}=(0,0,w)$  the integral form of Eq.\eqref{Eq:OmegDisc} reads
\begin{widetext}
\begin{eqnarray}
\nonumber\Omega(\beta,V,\mu,w,\mathcal{N})&=&-\frac{V
   (\mu+\mu_{0}) ^2}{2 g}+\int\limits_0^{P_{\rm max}}\int\limits_{-1}^1\frac{p^2 V}{2 \pi ^2   \hbar 
   ^3}\log{ \left(\beta\sqrt{\left(\mu+2\mu_{0}+\frac{p^2}{2 m} \right)^2-(\mu+\mu_{0})^2}-\beta pw z\right)}\frac{dz\,dp}{2}\hspace{1.5cm}\\
   &=&-\frac{V
   (\mu+2\mu_{0})}{2 g}-\frac{P_{\rm max}^3 V}{6 \pi ^2 \beta  \hbar ^3} \left\{\frac{2}{3}-\log{[\beta  \epsilon
   (P_{\rm max};\mu)]}-f\left[\frac{4 m \mu
   }{P_{\rm max}^2}\right]\left(1-\frac{w^2m}{2\mu}\right)-\frac{\mu_{0}}{\mu}f_{0}\left[\frac{4 m \mu
   }{P_{\rm max}^2}\right]\right\}\label{Eq:PartialOmegalowT}.
\end{eqnarray}
\end{widetext}
In order to obtain Eq.\eqref{Eq:PartialOmegalowT} the thermodynamic limit \eqref{Eq:ThermoLimit} of infinite volume \footnote{The thermodynamic limit is taken over the grand canonical potential $\Omega=-\beta^{-1}\log{\mathcal{Z}}$ as $ V\left(\displaystyle\lim_{V\to\infty} \frac{\Omega}{V}\right)$.} was taken and the conditions $w^2\ll\mu/m$, $\mu_{0}/\mu\ll1$ were used.  The functions $f[z]$ and $f_{0}[z]$ are explicited in Eqs.(\ref{Eq:f}-\ref{Eq:f0}). Note that the dependence of the grand canonical potential $\Omega$ on the number of modes $\mathcal{N}$ is implicitly given by $P_{\rm max}$ and Eq.(\ref{Eq:Nmodes}). The first term in $\Omega$ is due to the condensed mode at ${\bf p}=0$.

The low-temperature approximation to all thermodynamic functions is directly obtained from equation \eqref{Eq:PartialOmegalowT} by first setting $\mu_{0}=0$ and then differentiating \eqref{Eq:PartialOmegalowT}, using relation \eqref{Eq:dPartialOmega}.
It is straightforward to check that both definition of the pressure in Eq.\eqref{Eq:PresThermo} coincide. 
Furthermore the higher order moments of the density can be easily computed by taking successive derivatives of the grand canonical potential. For instance it is straightforward to show that the variance of the density $\rho$ (see solid line on histograms displayed on Fig.\ref{Fig:HisRho}.a) is given by
\begin{equation}
V^2\langle\delta\rho^2\rangle=-\beta^{-1}m^2\frac{\partial^2\Omega}{\partial \mu^2}.\label{Eq:deltarho}
\end{equation}

It can also be checked on the explicit expression for the entropy $S$ (see Eqs.\eqref{Eq:LowTFunctions}) that, as expected for a classical system, the entropy depends by a logarithmic term on the phase-space normalization. However the function $T S+ \lambda_\mathcal{N}\mathcal{N}$ is independent of phase-space normalization (see discussion below Eq.\eqref{Eq:ThermoRelation}).  

Finally, low-temperature expressions for the energies (\ref{Eq:defEkin}-\ref{Eq:defEq}) and their corresponding spectra can be easily obtained using Madelung's transformation \eqref{Eq:defMadelung}. At low temperatures the fluctuations are smalls and $e_{\rm kin}$ depends only on $\phi$ and $e_{\rm q}+e_{\rm int}$ only on $\rho$. The total energy is thus decomposed in two non-interacting terms. 
Equipartition of energy between the total kinetic energy $e_{\rm kin}$ and quantum plus internal energy $e_{\rm q}+e_{\rm int}$ is thus expected at low temperature.

The next subsections will be concerned with the vanishing counterflow case  $w=0$. The states with non-zero counterflow $w$ will be studied in details in section \ref{Sec:Meta}.

\subsection{$\lambda$ transition and vortices\label{SubSection:Lambda-Trans}}

To characterize the condensation transition, we present here four temperature scans performed using SGLE \eqref{Eq:SGLRphys}. Three of them are at resolution of $64^3$ with respectively constant chemical potential, density and pressure (using Eqs.(\ref{Eq:muAtRho}-\ref{Eq:muAtPres})). The fourth scan is performed at constant pressure but at a resolution of $128^3$. 
The coherence length is fixed so that $\xi \kmax=1.48$ is kept constant.

\begin{figure}[htbp]
\begin{center}
\includegraphics[height=8.3cm]{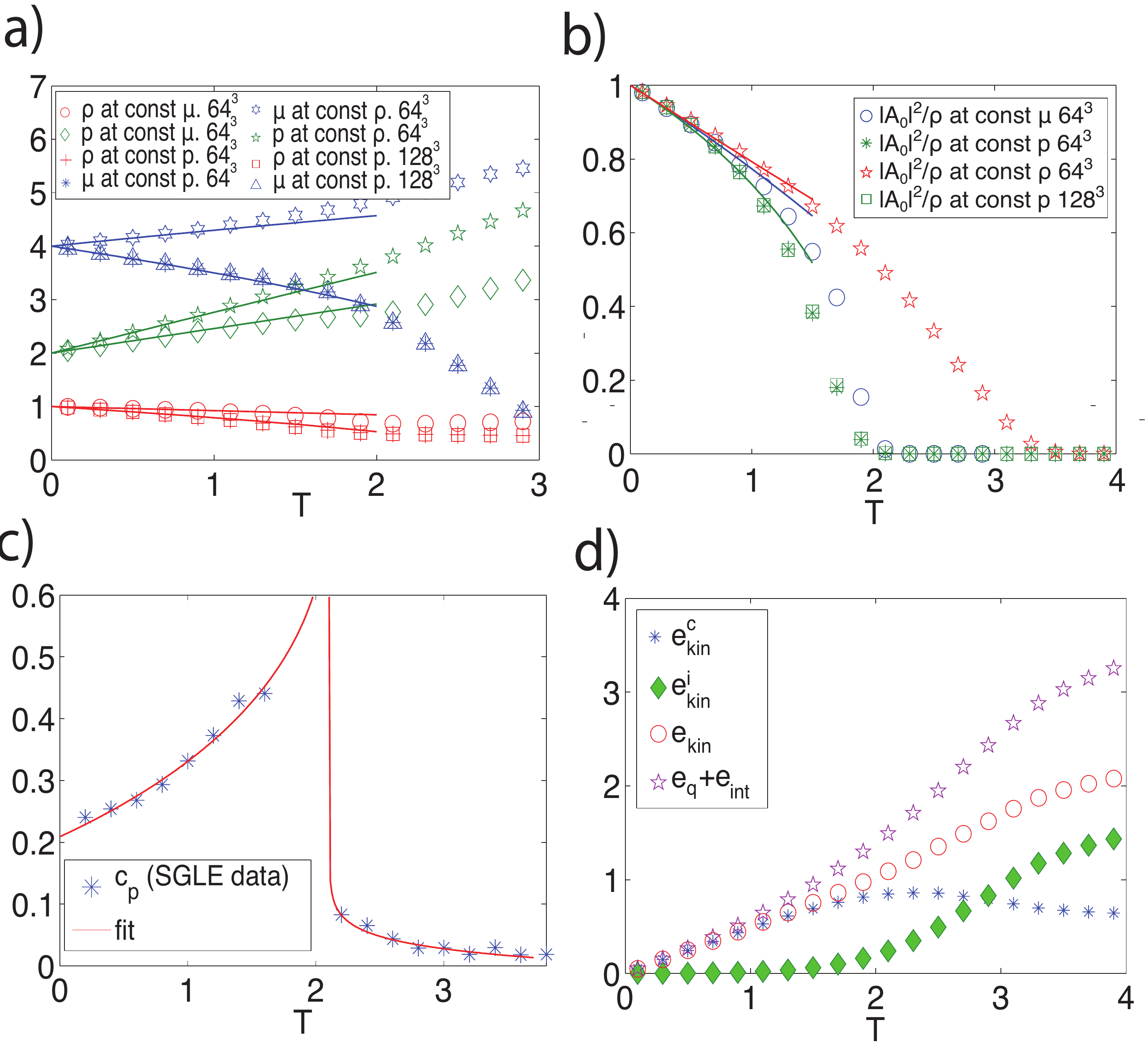}
\caption{(Color online) a) Temperature dependence of the density $\rho$, pressure $p$ and chemical potential $\mu$ for SGLE scans at constant density, pressure and chemical potential (see legend on figure). b) Temperature dependence of the condensate fraction $|A_{\bf 0}|^2/\rho$ (same scans as in a)). c) Specific heat $c_{p}=\left.\frac{\partial H}{\partial T}\right|_{p}$ at constant pressure and resolution $128^3$ the solid line corresponds to a fit (see Eq.\eqref{Eq:cpRG}). d) Temperature dependence of the energies $e_{\rm kin}^{\rm c}$,  $e_{\rm kin}^{\rm i}$,  $e_{\rm kin}$ and  $e_{\rm q}+e_{\rm int}$ at constant density; equipartition of energy between $e_{\rm kin}$ and  $e_{\rm q}+e_{\rm int}$ is apparent at low temperatures.}
\label{Fig:Scan}
\end{center}
\end{figure}

Figure \ref{Fig:Scan}.a displays the results of the scans. Observe that the low-temperature behavior is in good agreement with the analytical calculations of section \ref{sec:LowTcalulations} and the explicit formulae given in appendix \ref{App:Low-T}. Also observe that the constant pressure scans at resolutions of $64^3$ and $128^2$ coincide for all temperatures showing that the thermodynamic limit \eqref{Eq:ThermoLimit} discussed in section  \ref{SubSec:Thermo} is obtained at these resolutions.

Figure \ref{Fig:Scan}.b displays the temperatures dependence of the condensate fraction $|A_{\bf 0}|/\rho$ for the four SGLE runs. Observe that the condensation transition previously obtained (in the constant density case) by microcanonical simulations in references \cite{Davis:2001p1475,Connaughton:2005p1744} is reproduced and also present (at different critical temperatures) in the constant pressure and chemical potential scans. 

The SGLE algorithm directly provides the temperature as a control variable. It thus allows to easily obtain the specific heat from the data. Figure \ref{Fig:Scan}.c displays the specific heat at constant pressure $c_{p}=\left.\frac{\partial H}{\partial T}\right|_{p}$ for the scan at resolution $128^3$.
Let us remark that the ($w=0$) statistic weight of distribution (\ref{Eq:StatProb},\ref{Eq:defF}) corresponds to that of (standard two-component) second order phase transitions \cite{JAmit:1978p4003,ZinnJustin:2007p3931}. We thus expect the condensation transition visible on Fig.\ref{Fig:Scan}.c to be in this standard class.
This point is confirmed by the solid lines in Figure \ref{Fig:Scan}.c that correspond to a fit with the theoretical prediction given by the renormalization group (RG)
\begin{equation}
c_{p}=\frac{A^\pm}{\alpha_{\rm RG}}|\tau|^{-\alpha_{\rm RG}}(1+a_{c}^\pm|\tau|^{\Delta}+b_{c}^\pm |\tau|^{2\Delta}+\ldots)+B^\pm\label{Eq:cpRG}
\end{equation}
where $\tau=\frac{T-T_\lambda}{T_\lambda}$ and the $+$ and $-$ signs refer to $T>T_{\lambda}$ and $T<T_{\lambda}$, see reference  \cite{Lipa:2003p2679}. The fit was obtained in the following way: first the identification of the transition temperature $T_{\lambda}$ was done by finding the zero of the linear interpolation of the second order difference of $H$, discarding the three closest point to the zero of $|A_{\bf 0}|^2/\rho$. 
Then, using the critical exponents $\alpha_{\rm RG}=-0.01126$ and $\Delta=0.529$ given by the RG, the data was fitted as in reference \cite{Lipa:2003p2679} over the non-universal constant. The obtained values are $A^+/A^-=1.42$, to be compared with $1.05$ in reference \cite{Lipa:2003p2679}. This discrepancy is probably due to finite size effects.

Finally on Fig.\ref{Fig:Scan}.d the temperature dependence at constant density of the different energies (\ref{Eq:defEkin}-\ref{Eq:defEq}) expressed in terms of hydrodynamical variables is displayed. Observe that the incompressible kinetic energy $E_{\rm kin}^{\rm i}$ vanishes for low temperatures $T\ll T_{\lambda}^\rho$, where $T_{\lambda}^\rho=3.31$ is the transition temperature at constant density. This vanishing is connected to the disappearance of vortices, that it is also manifest in the density histograms in Fig.\ref{Fig:HisRho}. At low temperature equipartition of energy between the total kinetic energy $e_{\rm kin}$ and quantum plus internal energy $e_{\rm q}+e_{\rm int}$ (see the discussion at the end of section \ref{sec:LowTcalulations}) is apparent on Fig.\ref{Fig:Scan}.d.

\section{Energy cascade, partial thermalization and vortex annihilation\label{Sec:Cascade}}

A new mechanism of thermalization through a direct cascade of energy is studied in section \ref{SubSec:PartTherm}.
Using initial conditions with mass and energy distributed at large scales, a long transient with partial thermalization of the density waves is obtained at small-scales. Vortex annihilation is observed to take place and is related to mutual friction effects. 
A bottleneck effect that
produces spontaneous self truncation with partial thermalization and a time-evolving effective truncation wavenumber is
characterized in section \ref{SubSection:SelfTrunc} for large dispersive effects at the maximum wavenumber of the simulation.

\subsection{Partial thermalization\label{SubSec:PartTherm}}

We now study the (partial) thermalization of the superfluid Taylor-Green (TG) vortex. 
This flow, that was first introduced in reference \cite{Nore:1997p1331}, develops from an initial condition that is prepared by a minimization procedure using the advected real Ginzburg-Landau equation (ARGLE) \cite{Nore:1997p1331}. The nodal lines of the initial condition $\psi_{\rm TG}$ are the vortex lines of the standard TG vortex and obeys all its symmetries. Numerical integrations are performed with a symmetric pseudo-spectral code, making use of the TG symmetries to speed up the computations and optimize memory use, as described in reference \cite{Nore:1997p1331}. 
We use the equivalent to $64^3$, $128^3$, $256^3$ and $512^3$ collocation points and the coherence length is set such that, in all cases, $\xi \kmax=1.48$. 

\begin{figure}[htbp]
\begin{center}
\includegraphics[height=12.5cm]{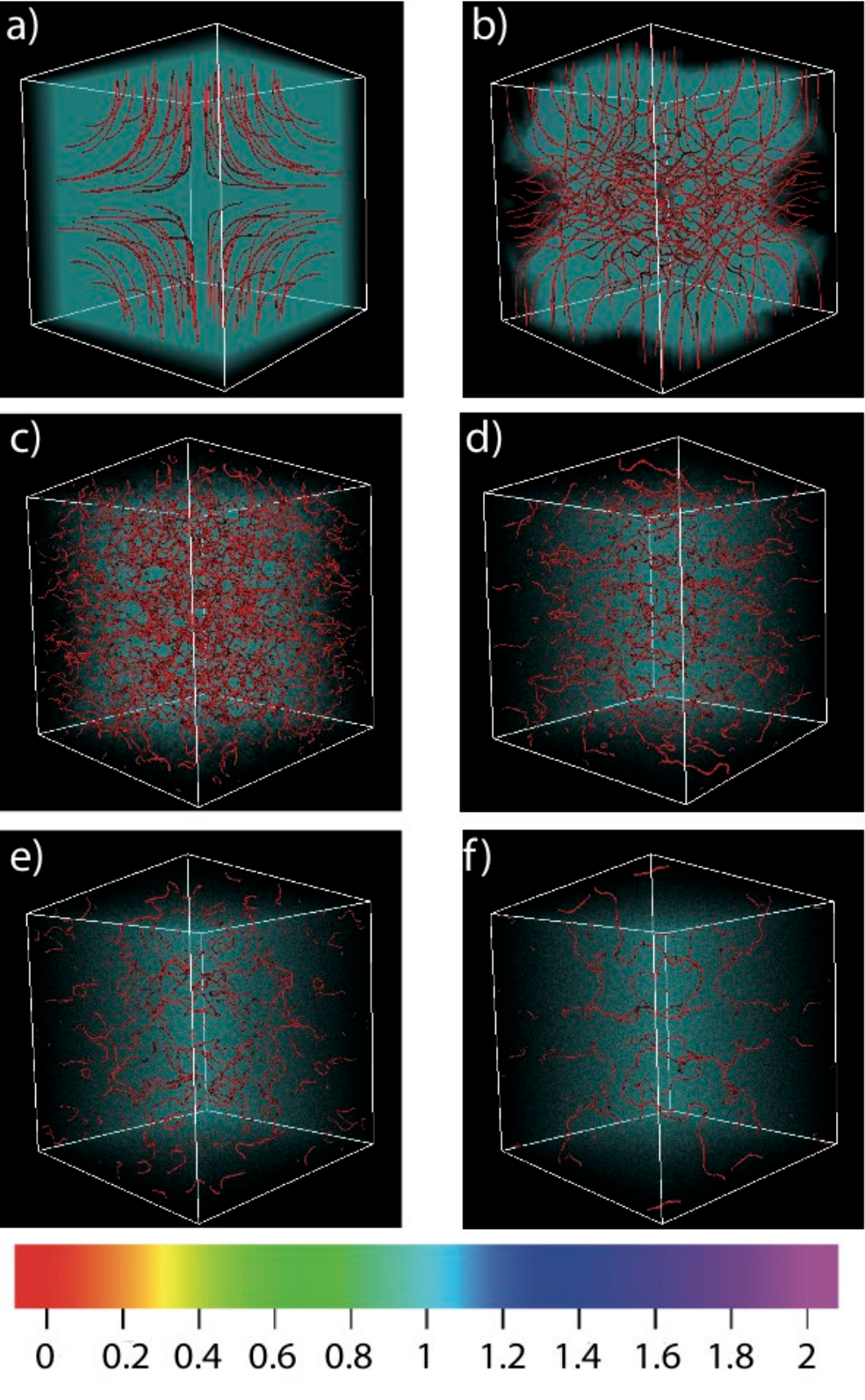}
\caption{(Color online) $3$D visualization of density at $t=0$, $5$, $10$, $20$, $31$ and $55$ at resolution $512^3$. Vortices are displayed as grey (red) isosurfaces and the grey (blue) clouds correspond to density fluctuations.}
\label{Fig:TG_Phys}
\end{center}
\end{figure}

Vortices and density fluctuations corresponding to the $512^3$ run are visualized on Fig.\ref{Fig:TG_Phys} using the VAPOR \footnote{http://www.vapor.ucar.edu} software. The short time behavior, see Fig.\ref{Fig:TG_Phys}.a-c, corresponds to the GPE superfluid turbulent regime previously studied in ref.\cite{Nore:1997p1333}. A new TGPE thermalization regime where vortices first reconnect into simpler structures and then decrease in size with the emergence of a thermal cloud is present at latter times, see Fig.\ref{Fig:TG_Phys}.d-e.

\begin{figure}[htbp]
\begin{center}
\includegraphics[width=.49\textwidth]{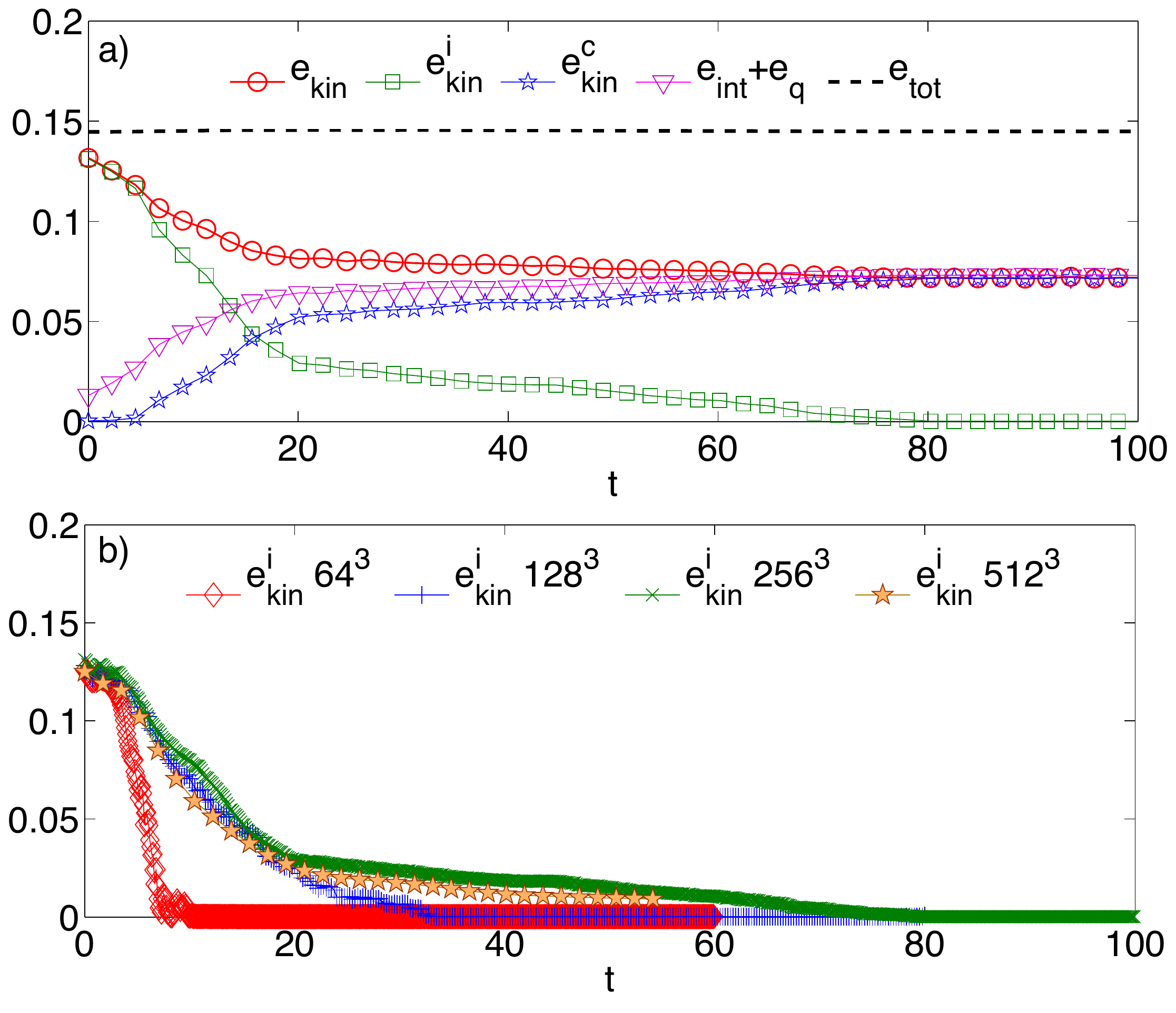}
\caption{(Color online) a) Temporal evolution of energies $e_{\rm kin}^{\rm c}$,  $e_{\rm kin}^{\rm i}$,  $e_{\rm kin}$ and  $e_{\rm q}+e_{\rm int}$ at resolution $256^3$.  At large times, the incompressible energy vanishes and equipartition of energy between $e_{\rm kin}$ and  $e_{\rm q}+e_{\rm int}$ is observed. b) Temporal evolution of $e_{\rm kin}^{\rm i}$ at resolution of $64^3$, $128^3$, $256^3$ and $512^3$ with constant $\xi \kmax=1.48$.}
\label{Fig:TGa}
\end{center}
\end{figure}

\begin{figure*}[htbp]
\begin{center}
\includegraphics[width=.9\textwidth]{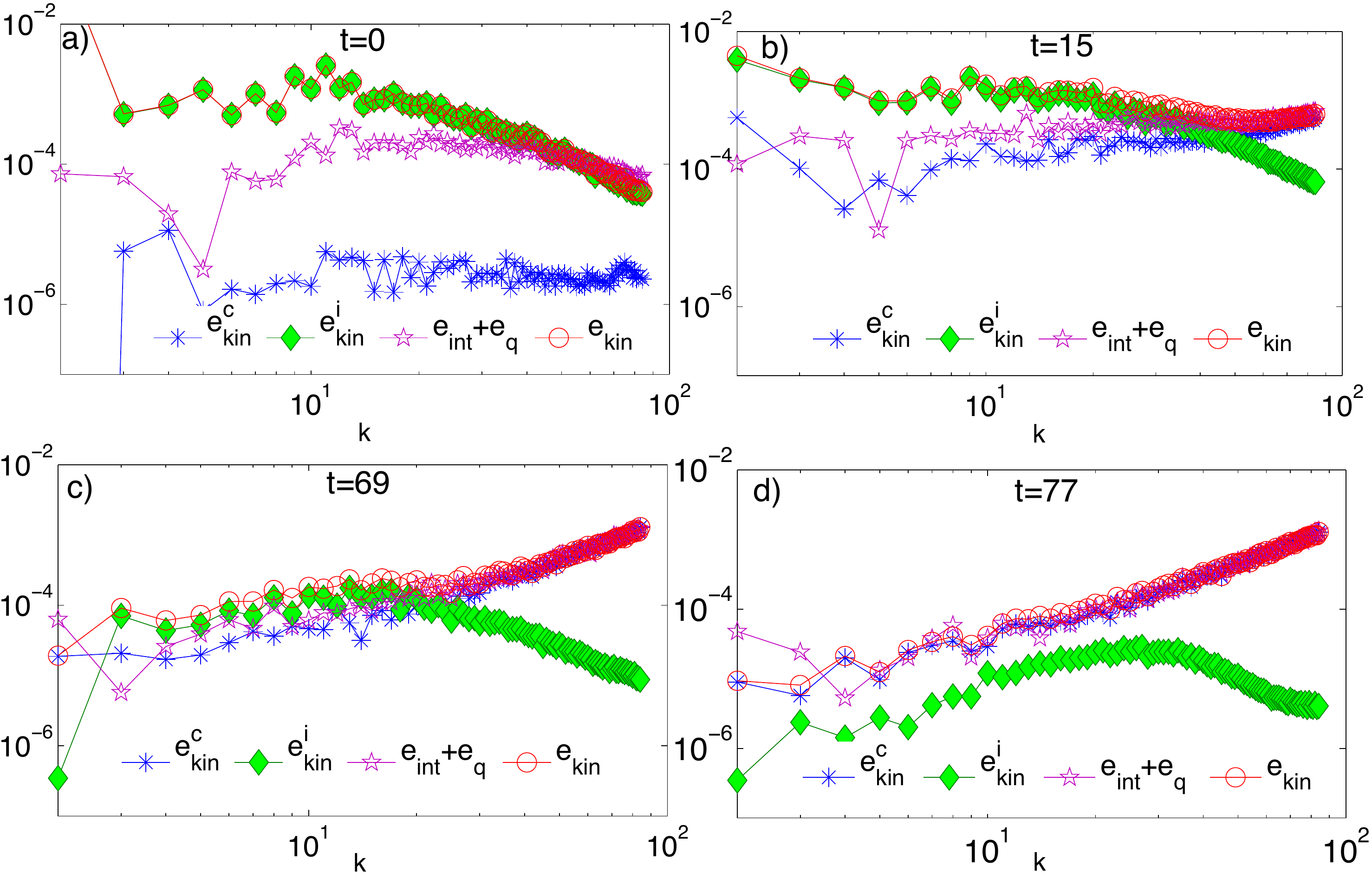}
\caption{(Color online) a-d) Energy spectra at $t=0$, $15$, $69$, $77$ at resolution $256^3$. Figure d) shows that equipartition is reached for all mode.}
\label{Fig:TGb}
\end{center}
\end{figure*}

To further study this new TGPE regime, the temporal evolution of $e_{\rm kin}$, $e_{\rm kin}^{\rm i}$, $e_{\rm kin}^{\rm c}$, $e_{\rm q}+e_{\rm int}$ is shown on Fig.\ref{Fig:TGa}.a and the corresponding energy spectra are displayed on Fig.\ref{Fig:TGb}. Observe that, at $t=0$, $e_{\rm kin}^{\rm i}$ contains almost all the energy because of the highly vortical initial condition.
The early times ($t\le 15$) behavior corresponds to the PDE regime of the GPE \eqref{Eq:GPE} that was previously reported in references \cite{Nore:1997p1333,Nore:1997p1331}. An energy transfer  is observed from $e_{\rm kin}^{\rm i}$ to the other energies ($e_{\rm kin}^{\rm c}$ and $e_{\rm int}+e_{\rm q}$) that are associated to the density waves. 

Continuing the temporal integration the spectral convergence to the GP partial differential equation is lost. The dynamics becomes influenced by the truncation wavenumbers $\kmax$ and thermalization starts to take place. Two new regimes are observed. The first one for $20\lesssim t\lesssim 80$ corresponds to a partial thermalization at small-scales: see Fig.\ref{Fig:TGb}.b-d. Observe that equipartition of $e_{\rm kin}^{\rm c}$ and $e_{\rm q}+e_{\rm int}$ begins to be established in this phase. The thermalized zone then progressively extends to larger wavenumbers. During this phase $e_{\rm kin}^{\rm i}$ decrease at almost constant rate (see Fig.\ref{Fig:TGa}.a). As shown on Fig.\ref{Fig:TGa}.b, this phase is delayed when the resolution is increased at constant $\xi\kmax$.

 Around $t=80$  (Fig.\ref{Fig:TGa}.a and \ref{Fig:TGb}.d) equipartition is established for each wave-number and $e_{\rm kin}^{\rm i}$ almost vanishes. 
 This vanishing is related to the disappearance of vortex lines that first reconnect into simpler structures which then decrease in size and number and finally disappear (as can be directly observed on the density visualizations corresponding to the $256^3$ run, pictures not shown). Note that the annihilation of the vortices can also be related to the contraction of vortex rings due to mutual friction reported in ref.\cite{Berloff:2007p423}. For $t>80$ the system finally reaches the thermodynamic equilibrium. The final absence of vortices and equipartition of energy between $e_{\rm kin}^{\rm c}$ and $e_{\rm q}+e_{\rm int}$, as can be directly checked on the temperature scan in Fig.\ref{Fig:Scan}.d, is a consequence of the low energy of the initial condition $\psi_{\rm TG}$. 
 
 We have thus presented for the first time a new mechanism of thermalization through a direct cascade of energy of the TGPE similar to that of the incompressible truncated Euler equation reported in reference \cite{Cichowlas:2005p1852}.

\subsection{Dispersive slowdown of thermalization and bottleneck\label{SubSection:SelfTrunc}}

We now turn to the study of dispersion effects on the thermalization of the TGPE dynamics and on vortex annihilation. To wit, we prepare three different initial conditions with different values of $\xi\kmax$ using the TG initial condition described in the preceding section. We fix the value of the coherence length to $\xi=\sqrt{2}/20$ and use resolutions of $64^3$, $128^3$ and $256^3$ corresponding to $\xi \kmax=1.48$, $2.97$ and $6.01$ respectively. The three initial condition therefore represent the same field at different resolutions.

\begin{figure}[htbp]
\begin{center}
\includegraphics[height=9cm]{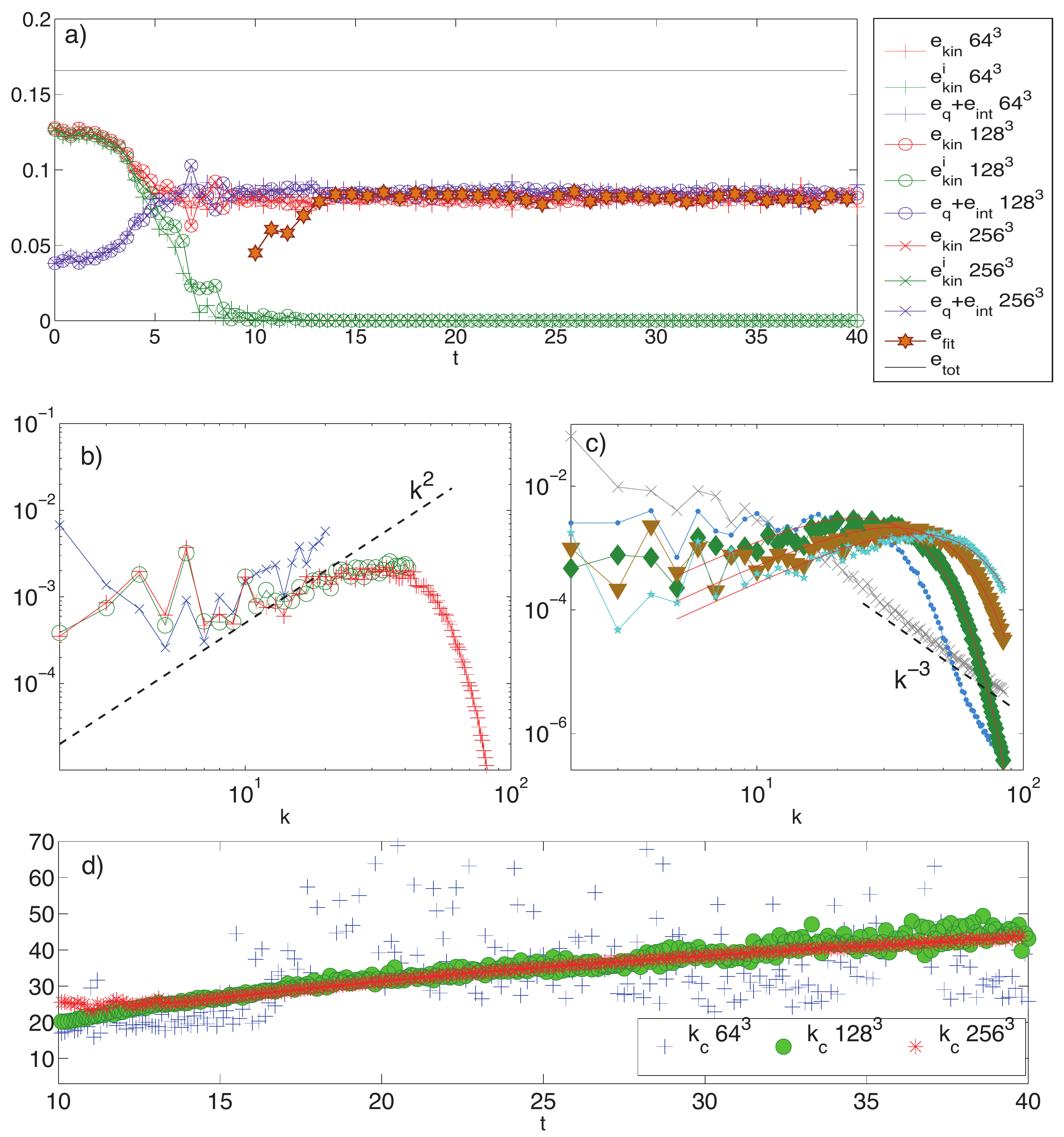}
\caption{(Color online) a) Temporal evolution of energies (as in Fig\ref{Fig:TGa}.a) for  $\xi\kmax=1.48$, $2.97$ and $6.01$ (resolution $64^2$, $128^3$ and $256^3$ respectively). Yellow stars are the kinetic energy reconstructed from fit data using Eq.(\ref{Eq:Ekinfit}). b) Kinetic energy spectrum at $t=17.4$ for $\xi\kmax=1.48$, $2.97$ and $6.01$; the dashed black line indicates $k^2$ power-law scaling. c) Temporal evolution of kinetic energy spectrum; the solid red lines correspond to fits using Eq.(\ref{Eq:spEkinfit}) and the dashes black line indicate $k^{-3}$ power-law scaling. d) Temporal evolution of effective self-truncation wavenumber $k_{c}$ (Eq.\eqref{Eq:spEkinfit}) at different resolutions.}
\label{Fig:SelfTrunc}
\end{center}
\end{figure}

The temporal evolutions of $e_{\rm kin}$, $e_{\rm kin}^{\rm i}$, $e_{\rm kin}^{\rm c}$ and $e_{\rm q}+e_{\rm int}$ for the three runs (indexed by the resolution) are displayed on Fig.\ref{Fig:SelfTrunc}.a. 
They are identical until $t\approx5$ where the run of resolution $64^3$ starts to lose its spectral convergence.
At at $t\approx20$ all runs appear to have thermalized on Fig.\ref{Fig:SelfTrunc}.a. However the kinetic energy spectra on Fig.\ref{Fig:SelfTrunc}.b shows a clear difference between the runs (the dashed line corresponds to $k^2$ power-law scaling).  The high-wavenumber modes of the $64^3$ run are thermalized. For the $128^3$ run the high-wavenumbers begin to fall down and, at resolution $256^3$, two zones are clearly distinguished. An intermediate thermalized range with an approximative $k^2$ power-law scaling is followed by a steep decay zone well before $\kmax=85$. Remark that in the $256^3$ run the spectral convergence is still ensured and the (partial) thermalization is thus obtained within the GP PDE-dynamics.

The temporal evolution of $e_{\rm kin}(k)$ for the $256^3$ run is displayed in Fig.\ref{Fig:SelfTrunc}.c. The large wave-number $k^{-3}$ power-law behavior at $t=0$ is an artifact of the high-$k$ decomposition of energies in the presence of vortices (see pp. 2649-2650 of ref.\cite{Nore:1997p1331} and \cite{krstulovic-2009}) and a faster decay is recovered as soon as the vortices disappear.
The thermalized intermediate zone is observed to slowly extends to smaller wave-numbers. This naturally defines a \emph{self-truncation} wave-number $k_{c}(t)$ where the energy spectrum starts to drastically decrease.

In order to determine $k_{c}(t)$ we have tested fits to $e_{\rm kin}(k)$ using two type of trial spectra with three free parameters: $e_{\rm fit \,I}(k)=A(t) k^{-n}\exp{[-2\delta(t)k]}$ and $e_{\rm fit \,II}(k)=A(t) k^{-n}\exp{[-\gamma(t)k^2]}$. The $e_{\rm fit \,II}(k)$ fit was found to work better in the sense that it both gives the correct  $n=-2$ prefactor at intermediate and large times and also gives a better fit to the data at high $k$ (data not shown). 
Fixing the prefactor at the value $n=-2$, we finally define our working two-parameter fit as:
\begin{eqnarray}
e_{\rm fit}(k,t)&=&A(t) k^2 e^{- \left[\left({\frac{9 \pi}{16}}\right)^{\frac{1}{3}}\left({\frac{k}{k_c(t)}}\right)^{2}\right]}\label{Eq:spEkinfit}\\
e_{\rm fit}(t)&=&\int_0^{k_{\rm max}}e_{\rm fit}(k,t) dk.\label{Eq:Ekinfit}
\end{eqnarray}
The factor $(9\pi/16)^{1/3}$ in Eq.\eqref{Eq:spEkinfit} was set in order to obtain the limits $Ak_{c}^3/3$ and $A\kmax^3/3$ for $e_{\rm fit}(t)$ when $\kmax\to\infty$ and $k_{c}\to\infty$ respectively. The fits are also displayed in Fig.\ref{Fig:SelfTrunc}.c. They are in good agreement with the data after vortices have disappeared. The temporal evolution of $e_{\rm fit}(t)$ is displayed in Fig.\ref{Fig:SelfTrunc}.a. It does converge to the thermalized value of the energy.
Finally the temporal evolution of the self-truncation wavenumber $k_c(t)$, which seems to have a well defined limit at infinite resolution, is displayed in Fig.\ref{Fig:SelfTrunc}.d for the three runs. 

An open question is wether $k_c$ is bounded in time in the PDE regime where $k_c\ll k_{\rm max}$.
In other words, is thermalization of the $\xi k_{\rm max}\gg1$ truncated system simply delayed or completely inhibited when $\xi k_{\rm max}$ is large enough?
Note that this problem is related to the classical Fermi-Pasta-Ulam-Tsingu problem \cite{Fermi:1955p3871}.

To try to answer this question within the Taylor-Green framework would be computationally very expensive as long runs should be performed at arbitrarily high resolutions.

\begin{figure*}[htbp]
\begin{center}
\includegraphics[width=1.\textwidth]{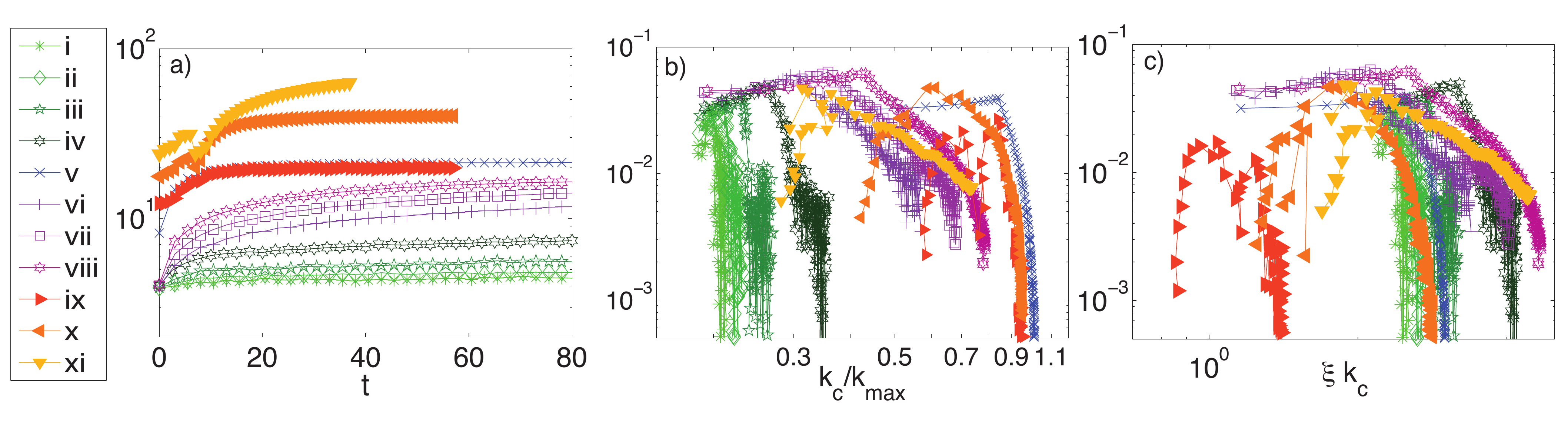}
\caption{(Color online) a) Evolution of the self-truncation wavenumber $k_c$. Curves  i-iv: $\xi=2\sqrt{2}/5$, $k_c^{\rm in}=4$, $e^{\rm in}=0.1,\,0.2\,,0.4\,,1$; v: $\xi=\sqrt{2}/10$, $k_c^{\rm in}=8$, $e^{\rm in}=0.2$; vi-viii: $\xi=\sqrt{2}/5$,  $e^{\rm in}=0.1,\,0.2\,,0.4$ (i-viii in resolution $64^3$); ix-xi: Taylor-Green resolutions $64^3$, $128^3$ and $256^3$.  b-c) Parametric representation $dk_{c}/dt$ v.s. $k_{c}/\kmax$ and $dk_{c}/dt$ v.s. $\xi k_{c}$ (same labels as in Fig.a).}
\label{Fig:SelfTrunc2}
\end{center}
\end{figure*}

A simple alternative idea to study this problem is to use initial data for the TGPE generated by the SGLE with a variable truncation wavenumber $k_c^{\rm in}$, set to a target value of $k_c$, smaller than the maximum truncation wavenumber $k_{\rm max}$ allowed by the resolution. 
This SLGE-generated initial data can then be used to run the TGPE at a given value of $\xi k_c$ with arbitrarily large values of $\xi k_{\rm max}$. A number of runs were performed at resolution $64^3$ with various values of $k_c^{\rm in}$, $\xi$, and initial energy $e^{\rm in}$ (see legend on Fig.\ref{Fig:SelfTrunc2}). 
The result of these computations are compared with the above Taylor-Green runs (see Fig.\ref{Fig:SelfTrunc}) and displayed on Fig.\ref{Fig:SelfTrunc2}. Because of the steep decay of the energy spectrum for $k \gg k_c$, the self-truncation wavenumber is determined using the integral formula
\begin{equation}
k_c=\sqrt{\frac{5}{3}\frac{\int_0^{\kmax} k^2e_{\rm kin}(k)dk}{\int_0^{\kmax} e_{\rm kin}(k)dk}}\label{Eq:kcInt}.
\end{equation}

A general growth in time of $k_c$ is apparent on Fig.\ref{Fig:SelfTrunc2}.a for both the Taylor-Green runs and the SGLE-generated initial data, showing similar behavior. In order to check for self-similar regime a parametric Log-Log representation 
$dk_{c}/dt$ v.s. $k_{c}$ has been used on Fig.\ref{Fig:SelfTrunc2}.b and Fig.\ref{Fig:SelfTrunc2}.c. With this representation, a self-similar evolution $k_c(t)\sim t^\eta$ corresponds to a line of slope $\chi=(\eta-1)/\eta$.  Figure \ref{Fig:SelfTrunc2}.b,  shows transient self-similar evolutions, that all terminate by a vertical asymptote, corresponding to logarithmic growth ($\eta=0$). This self-truncation takes place for small values of $k_{\rm c}/\kmax$ strongly suggesting that the self-truncation happens in a regime independent of cut-off. Finally, 
Fig.\ref{Fig:SelfTrunc2}.c suggests that, depending on initial conditions, self-truncation can take place at arbitrarily values of $\xi k_c$.

As the dynamics of modes at wave-numbers larger than $k_{c}$ is weakly nonlinear, it should be amenable to a description in terms of wave turbulence theory; this could perhaps explain the slowdown of the thermalization in this zone. The new regime indicates that total thermalization is delayed when increasing the amount of dispersion (controlled by $\xi k_{\rm max}$) but is preceded by a partial thermalization (quasi-equilibrium up to $k_{c}$)  \emph{within}  a PDE.

We now turn to estimations of order of magnitude relevant to physical BEC. At low-temperature, the GPE is known \cite{Proukakis:2008p1821} to give an accurate description of the (classical) dynamics of physical BEC at scales larger than the interatomic separation $\ell$. 
At finite temperature the TGPE gives a good approximation of Bose-Einstein condensate (BEC) only for the phonon modes with high occupation number, see \cite{Davis:2001p1475,Proukakis:2008p1821}.
At very low temperature thus only a limited range of low-wavenumber density waves are in equipartition.

This limited range has consequences on the low-temperature thermodynamics of BEC that can be obtained by the following considerations. 
The equipartition range is determined by the relation $k \le k_{\rm eq}$ with $\hbar \omega_{\rm B}(k_{\rm eq})=k_{\rm B}T$ where the Bogoluibov dispersion relation $\omega_{\rm B}(k)$ is given by
\begin{equation}
\omega_{\rm B}(k)=k\sqrt{\frac{g|A_{\bf 0}|^2}{m}+\frac{\hbar^2}{4m}k^2}.\label{Eq:bogu}
\end{equation}
The coherence length $\xi$ defined in Eq.\eqref{Eq:defxi} can be expressed in terms of the $s$-wave scattering length $\tilde{a}$ defined by $g=4 \pi  \tilde{a} \hbar^2 /m$ and the mean inter-atomic particle distance $\ell\equiv n^{-1/3}\approx|A_{0}|^{-2/3}$ as
\begin{equation}
\xi=(8\pi n \tilde{a})^{-1/2}=\ell \frac{1}{\sqrt{8\pi}}\left(\frac{\ell}{\tilde{a}}\right)^{1/2}.\label{Eq:xiBEC}
\end{equation}
For weakly interacting BEC the coherence length thus satisfies $\xi\gg\ell$. The equipartition wavenumber $k_{\rm eq}$ explicitly reads
\begin{equation}
k_{\rm eq}=\left[\frac{\sqrt{32 k_{\rm B}^2 m^2 T^2 \xi ^4+\hbar ^4}}{4 \xi ^2 \hbar
   ^2}-\frac{1}{4 \xi ^2}\right]^{1/2}.
\end{equation}

Using the Bose-Einstein condensation temperature of non-interacting particles (valid for $\tilde{a}\ll\ell$) \cite{Proukakis:2008p1821}
\begin{equation}
T_\lambda=\frac{2\pi\hbar^2}{k_{\rm B} m}\left[\frac{n}{\zeta(\frac{3}{2})}\right]^{2/3}\label{Eq:Tlambda}
\end{equation}
where $\zeta(3/2)=2.6124\ldots$, the equipartition wavenumber $k_{\rm eq}$ can be expressed as
\begin{equation}
k_{\rm eq}=\frac{1}{2\xi}      \left[ \sqrt{1+\frac{128\pi^2}{\zeta(\frac{3}{2})^{4/3}}   \frac{\xi^4}{\ell^4}   \frac{T^2}{T_\lambda^2}}-1 \right]^{1/2}\label{Eq:TstarBEC}
\end{equation}
Observe that $k_{\rm eq}$ varies from $k_{\rm eq}=0$ at $T=0$ to wavenumber of order $k_{\rm eq}\sim \ell^{-1}$ at $T_\lambda$ and it is equal to $k_\xi=2\pi/\xi$ at $T^*$ defined by
\begin{equation}
T^*=4 \pi  \sqrt{1+8 \pi ^2}  \zeta(\frac{3}{2}) ^{2/3}\frac{\tilde{a}}{\ell}T_\lambda.\label{Eq:Tstar}
\end{equation}
Thus the thermodynamic of physical BEC at low-temperature (e.g. specific heat scaling as $T^3$) can be recovered from the TGPE thermodynamics by setting $\kmax=k_{\rm eq}(T)$.

In experimental turbulent weakly interacting BEC such as \cite{Henn:2009p5720}  the value of $\xi k_{\rm eq}$ is large because $T^*/T_\lambda\sim \frac{\tilde{a}}{\ell}\ll1$ and therefore the corresponding TGPE should have a large $\xi\kmax$.
Thus the thermalization slowdown caused by the dispersive bottleneck should be in principle present in physical BEC, unless it is overwhelmed by other relaxation mechanisms \cite{Krstulovic:2010p5966}.  

\section{Metastability of counterflow, mutual friction and Kelvin waves\label{Sec:Meta}}

Counter-flow states with non-zero values of momentum generated by the new SGLE algorithm and their interaction with vortices are investigated in this section.
The counter-flow states are shown to be metastable under SGLE evolution; the spontaneous nucleation of vortex ring and the corresponding Arrhenius law are characterized in section \ref{subsec:Meta}.
Dynamical counter-flow effects are investigated in section \ref{subsec:MetaNLS} using vortex rings and vortex lines patterns
that are exact solutions of the GPE. 
Longitudinal and transverse mutual friction effects are produced and measured. An anomalous translational velocity of vortex
ring is exhibited and is quantitatively related to the effect of thermally excited finite-amplitude Kelvin waves. 
Orders of magnitude are estimated for the corresponding effects in BEC and superfluid $^4{\rm He}$.

\subsection{Metastability of grand canonical states with counterflow \label{subsec:Meta}}

\subsubsection{Thermodynamic limit of states with nonzero counterflow \label{subsubsec:TLMeta}}

The counterflow states with ${\bf W\neq0}$ are determined by thermal fluctuations around the minima of the energy
 $F$  Eq.\eqref{Eq:defF}. These minima correspond to the solution of
 \begin{equation}
 \fder{F}{\psi^*}=0=-\frac{\hbar^2}{2m} \gd \psi + \bet\mathcal{P}_{\rm G} [|\psi|^2]\psi-\mub \psi+i\hbar{\bf W}\cdot{\bf\nabla} \psi 
\end{equation}
that are plane-waves of the form
\begin{equation}
\psi(x; {\bf v}_{\rm s})=g^{-\frac{1}{2}}\sqrt{\mu-m{\bf W}\cdot  {\bf v}_{\rm s}+\frac{m v_{\rm s}^2}{2}}e^{-i\frac{m}{\hbar} {\bf v}_{\rm s}\cdot{\bf x}}, \label{Eq:psimeta}
\end{equation}
where the velocity $ {\bf v}_{\rm s}$ indexes the different solutions. 

In the thermodynamic limit, the Galilean group defined by the transformations (\ref{Eq:GaliTransf}-\ref{Eq:Ptrans}) is continuously indexed by the velocity $ {\bf v}_{\rm G}$. All wavefunctions \eqref{Eq:psimeta} are thus equivalent by Galilean transformation (and redefinition of the chemical potential). Under the Galilean transformation \eqref{Eq:GaliTransf} the energy $F$ is transformed as $F'=F-(m{\bf W}\cdot  {\bf v}_{\rm G}-m v_{\rm G}^2/2)N+{\bf v_{\rm G}}\cdot{\bf P}$. Note that, among all the minima of $F$ the one with $ {\bf v}_{\rm s}={\bf W}$ minimizes $F'$. This state corresponds to a condensate moving with uniform velocity ${\bf W}$. The ${\bf W}\cdot {\bf P}$ term is thus only imposing a Galilean transformation of the global minimum.

However, when working in a finite volume, the Galilean transformation is quantized (see Eq.\eqref{Eq:discValuesVs}). The minima of $F'$ of lowest energy then corresponds to a condensate moving with the quantized uniform velocity ${\bf v}_{\rm s}$ that is the closest to ${\bf W}$.
At finite temperature and volume, when ${\bf W}$ is not too large with respect to the velocity quantum in  Eq.\eqref{Eq:discValuesVs}, we have two ways to produce momentum in the system. 
The first one corresponds to Galilean transformations: ${\bf v}_{\rm s}\ne0$ in \eqref{Eq:psimeta}. The second one to fluctuations of the exited phonons, with ${\bf v}_{\rm s}=0$ in \eqref{Eq:psimeta} and the momentum of phonons imposed by the term ${\bf W}\cdot {\bf P}$ in the grand canonical distribution \eqref{Eq:StatProb}. 
Metastability is thus expected when ${\bf W}\ne 0$ with quasi-equilibrium corresponding to condensates at different wavenumbers with an energy barrier between each of those states.

In the context of the Landau two-fluid model \cite{Landau6Course} the velocity ${\bf v}_{\rm s}$ of the condensate corresponds to the superfluid velocity and the momentum carried by the exited phonons is written as ${\bf P}=\rho_{\rm n}({\bf v}_{\rm n}-{\bf v}_{\rm s})$ where $\rho_{\rm n}$ and ${\bf v}_{\rm n}$ are called the normal density and velocity respectively. The counterflow velocity defined by $\widetilde{{\bf W}}= {\bf v}_{\rm n}- {\bf v}_{\rm s}$ is a Galilean invariant.

The above discussion shows that, in general  the variable ${\bf W}$ in the SGLE \eqref{Eq:SGLRphys} corresponds to ${\bf W}={\bf v}_{\rm n}$. In the thermodynamic (infinite volume) limit ${\bf W}={\bf v}_{\rm s}$ and there is thus no counterflow $\widetilde{{\bf W}}={\bf v}_{\rm n}-{\bf v}_{\rm s}={\bf 0}$. For finite-size systems, in general ${\bf v}_{\rm s}\neq{\bf W}$ and $\widetilde{{\bf W}}\neq{\bf 0}$.

We thus define (when $v_{\rm s}=0$) the normal density by
\begin{equation}
\rho_{\rm n}=\left.\frac{\partial P_{z}}{\partial w_{z}}\right|_{w_{z}=0}.\label{Eq:defRhon}
\end{equation}

\subsubsection{Thermodynamics of metastable states at small temperature and small counterflow}

\begin{figure}[htbp]
\begin{center}
\includegraphics[height=8.1cm]{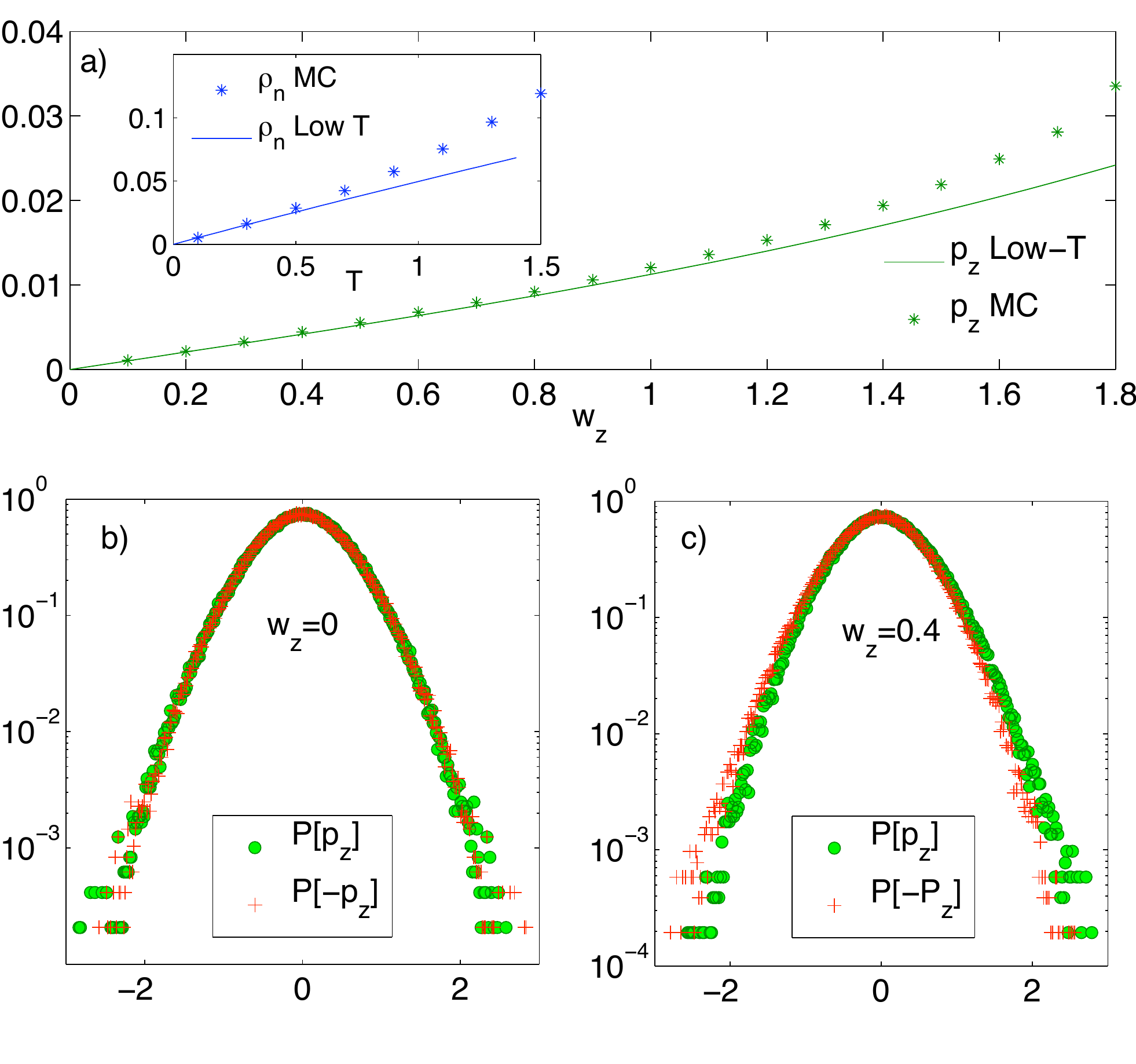}
\caption{(Color online) Counterflow dependence of momentum $P_{z}$ ($w_{x}=w_{y}=0$). Inset: Temperature dependence of $\rho_{\rm n}=\left.\frac{\partial P_{z}}{\partial w_{z}}\right|_{w_{z}=0}$. b) Histograms of momentum $P_{z}$ and $-P_{z}$ (in $log-lin$)  with no counterflow at $T=1$. No asymmetry is observed. c) Histograms of momentum $P_{z}$ and $-P_{z}$ with  counterflow $w_{z}=.4$ at $T=1$. An asymmetry, induced by counterflow, is apparent. Observe that both histograms are centered at $P_{z}=0$.}
\label{Fig:HistoPz}
\end{center}
\end{figure}

In order to first validate the SGLE in the presence of counterflow two scans are performed at constant density using a resolution of $64^3$ and $\xi\kmax=1.48$. The condensate is set at ${\bf k}=0$ in the SGLE initial data and the temperature is fixed to $T=0.2$. This low temperature allows us to increase the value of the counterflow $w_{z}$ (hereafter we set $w_{x}=w_{y}=0$) keeping the condensate at ${\bf k}=0$.  The dependence of the momentum $P_{z}$ on $w_{z}$ is presented in Fig.\ref{Fig:HistoPz}.a. 
The solid line corresponds to the low-temperature calculations (Eq.\eqref{Eq:PartialOmegalowT} and appendix \ref{App:Low-T} Eqs.\eqref{Eq:LowTFunctions}). The second run correspond to a temperature scan (at low counterflow $w_{z}=.1$). The temperature dependence of $\rho_{\rm n}$ is displayed together with the low-temperature calculation on the inset of Fig.\ref{Fig:HistoPz}.a.

Figure \ref{Fig:HistoPz}.a-b display histograms of $P_{z}$ and $-P_{z}$ in physical space, both obtained at $T=1$ with the condensate at ${\bf k}=0$ but with zero and non-zero counterflow. Observe that the histograms are both centered at $P_{z}=0$ but the non-zero counterflow induces an asymmetry in the statistical distribution that yields a non-zero value for the mean momentum.

\subsubsection{Spontaneous nucleation of vortex rings and Arrhenius law}

At temperatures and counterflow velocities large enough the stochastic process defined by the SGLE can jump between the different metastable states discussed above in section \ref{subsubsec:TLMeta}. In this section, we show how the different states are explored, under SGLE evolution, by spontaneous nucleation of vortex rings.
To wit, we present a numerical integration of SGLE at resolution $64^{3}$ with $\xi\kmax=1.48$. With this choice of parameters the velocity quantum \eqref{Eq:discValuesVs} is fixed to $0.2$. The temperature is set to $T=0.775$ and the counterflow to $w_{z}=0.8$. The condensate is set at ${\bf k=0}$ in the SGLE initial data and the density is kept constant to $\rho=1$.

\begin{figure}[htbp]
\begin{center}
\includegraphics[height=9cm]{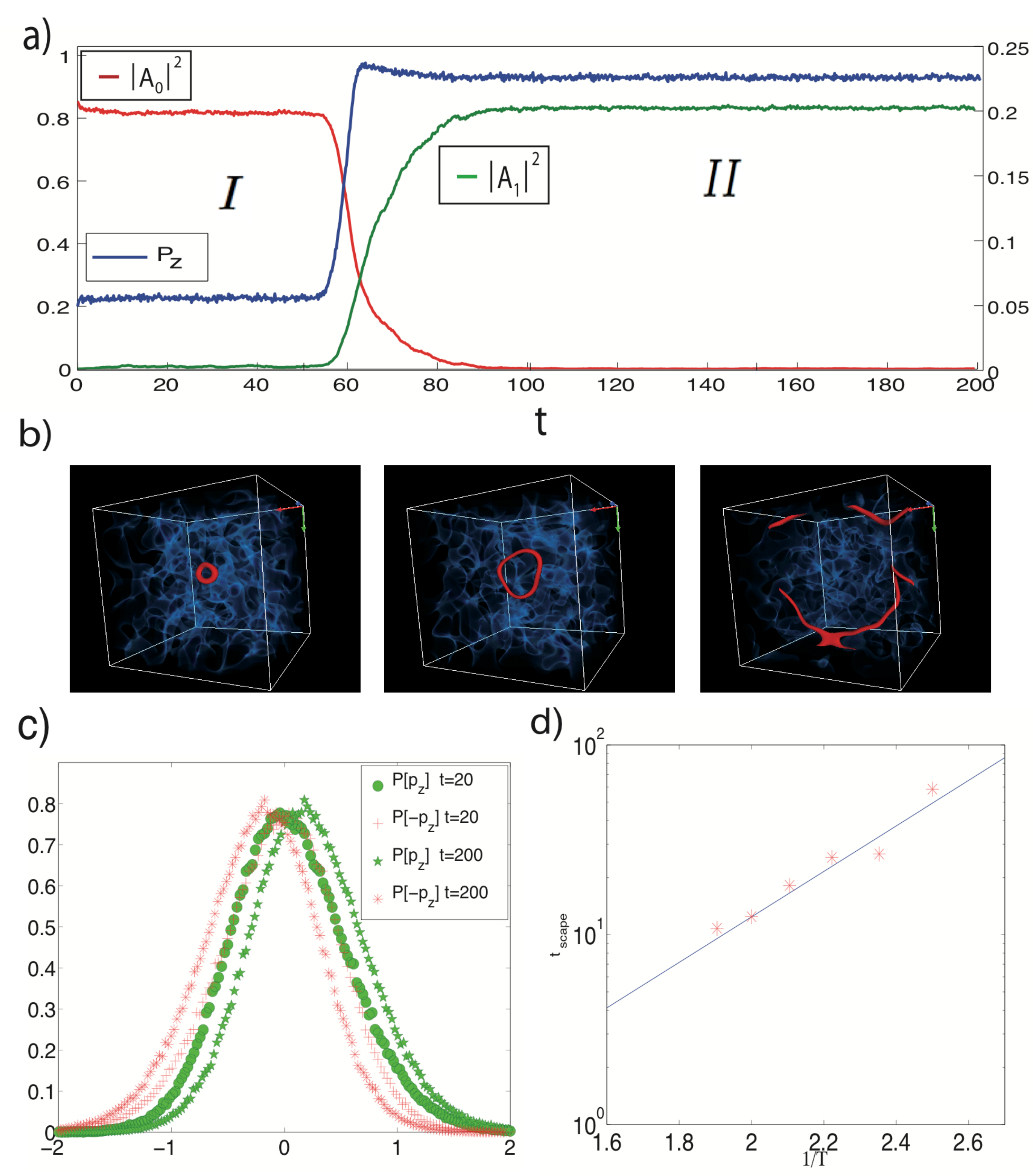}
\caption{(Color online) a) Temporal evolution of $|A_{\bf 0}|^2$ and $|A_{\bf 1}|^2$ (left scale) under SGL dynamics. Observe that there are two \emph{quasi}-stationary states ($I$ and $II$) and the condensate makes a transition from $k=0$ to $k=1$. The temporal evolution of the momentum $p_{z}$ is displayed in the same plot (right scale). Observe that transition from one state to the other is accompanied by an increase of momentum. b) $3$D visualization of density at $t=54.5$, $t=56$ and $t=60.5$;  the grey (blue) clouds corresponds to density fluctuations and the vortices are displayed as grey (red) isosurface (see colorbar on Fig.\ref{Fig:Kelvin} below). c) Histogram of momentum $p_{z}$ and $-p_{z}$ at the two \emph{quasi}-stationary states ($I$ and $II$) in $lin-lin$ plot. d) Arrhenius law: data form SGL dynamics (points) and theoretical Eq.\eqref{eqdef:Arrh} (solid line).}
\label{Fig:meta}
\end{center}
\end{figure}

The temporal evolution of the momentum $P_{z}$ is displayed in Fig.\ref{Fig:meta}.a (right scale). 
Observe that the system first spends some time at the state ($I$) with $P_{z}\approx 0.05$ and that, around $t=55$, it jumps to the state ($II$) with $P_{z}\approx 0.225$. These two metastable states correspond to quasi-equilibrium at ${\bf k=0}$ and ${\bf k=1}$ as is apparent in Fig.\ref{Fig:meta}.a (left scale) where the temporal evolution of $|A_{\bf 0}|^{2}$ and $|A_{\bf 1}|^{2}$ (see Eq.\eqref{eq:defFour}) are displayed.

In order to illustrate the dynamic of the condensate jump from ${\bf k=0}$ to ${\bf k=1}$ we now present $3$D visualization of the density at $t=54.5$, $t=56$ and $t=60.5$ on Fig.\ref{Fig:meta}.b. To produce this figure, the wave-function $\psi$ was first low-pass filtered and the density was then visualized using the VAPOR software. 
At early times ($t<50$, pictures not shown) no vortices are present in the box. At $t\approx54$ a vortex ring is nucleated. It then increases its size under SGLE evolution until it reconnects with the neighbor rings (recall that periodic boundary condition are used).  The ring finally contracts and disappears (pictures not shown). During this evolution, the local phase defect of the ring becomes global and changes the condensate wavenumber. Histograms of momentum $P_{z}$ and $-P_{z}$ in the two metastable states $I$ and $II$ are presented on Fig.\ref{Fig:meta}.c.  Observe that both momentum histograms of metastable states are asymmetrical (as it was the case on Fig.\ref{Fig:HistoPz}.c). However, note that $I$ is centered at $P_{z}=0$ and $II$ at $P_{z}=0.2$, respectively corresponding to the wavenumbers ${\bf k=0}$ and ${\bf k=1}$.

It is well known that the escape time of a metastable quasiequilibirum is given, in general, by an Arrhenius law \cite{Huepe:1999p1467,Gardiner1996Handbook}
\begin{equation}
t_{\rm esc}\sim t_{c}e^{-\beta \Delta F}\label{eqdef:Arrh},
\end{equation}
where $\Delta F$ is the activation energy of the nucleation solution and $t_{c}$ is a characteristic time. 
Here, the nucleation solution is given by a vortex ring that satisfies $\pder{F}{\psi^{*}}=0$. The energy barrier is thus determined by $\Delta F=H_{\rm ring}(R^*)-{\bf V}_{\rm ring}\cdot {\bf P}_{\rm ring}(R^*)$, where the analytic expressions for the energy $H_{\rm ring}$, the momentum ${\bf P}_{\rm ring}$ and the radius are given by
\begin{eqnarray}
V_{\rm ring}&=&\frac{\hbar}{2m}\frac{1}{R^*
}[\ln{(\frac{8 R^*
}{\xi})}-a  ]\label{Eq:vring}\\
P_{\rm ring}^*&=&\frac{2\pi^2\hbar\rho_\infty}{m}{R^*}
^2\label{Eq:Pring}\\
H_{\rm ring}^*&=&\frac{2 \pi^2\hbar^2}{m^2}\rho_{\infty}R^*
[\ln{(\frac{8 R^*
}{\xi})}-1-a  ]\label{Eq:Hring}
\end{eqnarray}
where $\rho_{\infty}$ is the density at the infinity and $a$ is a core model-depending constant with value $a=0.615$ for the GPE vortices \cite{Donne}. 
Formulae (\ref{Eq:vring}-\ref{Eq:Hring}) and the value of $a$ have been numerically validated in reference \cite{Winiecki:1999p607} using a Newton method \cite{Huepe:2000pp126-140,Tuckerman:2004Newton,Pham:2005Boundary}.

In order to numerically check that the escape time indeed follows an Arrhenius law we now perform runs with with $\xi\kmax=1.48$ and resolution  $32^{3}$. The counter-flow is fixed at $w=1.4$ and the condensate is set initially at ${\bf k=0}$ (constant density $\rho=1$). At each fixed temperature $T$, several numerical integration of SGLE are performed  and the escape times for the condensate to leave the wavenumber ${\bf k=0}$ are measured. These escape times are then averaged over more than $10$ realizations. Figure \ref{Fig:meta}.d displays the escape time $t_{\rm esc}$ obtained in this way as a function of the inverse temperature $1/T$ in $log-lin$. The slope of the solid line is computed using the analytic formulae (\ref{Eq:vring}-\ref{Eq:Hring}) of $\Delta F$. Both, numerical and theoretical Arrhenius laws are in good agreement.
The main consequence of this Arrhenius law is that it is practically possible to use the SGLE dynamics to prepare metastable states with finite value of counterflow and lifetime quantitatively given by \eqref{eqdef:Arrh}.

\subsection{Dynamical effects of finite temperature and counterflow on vortices \label{subsec:MetaNLS}}

We now turn to the study the dynamical effects of counterflow on TGPE vortex evolution. 
To wit, we set up finite temperature and finite counterflow initial states that also contain vortices. 
Two cases are investigated: (i) vortex lines, in a crystal-like pattern that does not produce self induced velocity and (ii) vortex rings, producing self induced velocity.

\subsubsection{Lattice of vortex lines}

\begin{figure}[htbp]
\begin{center}
\includegraphics[height=10cm]{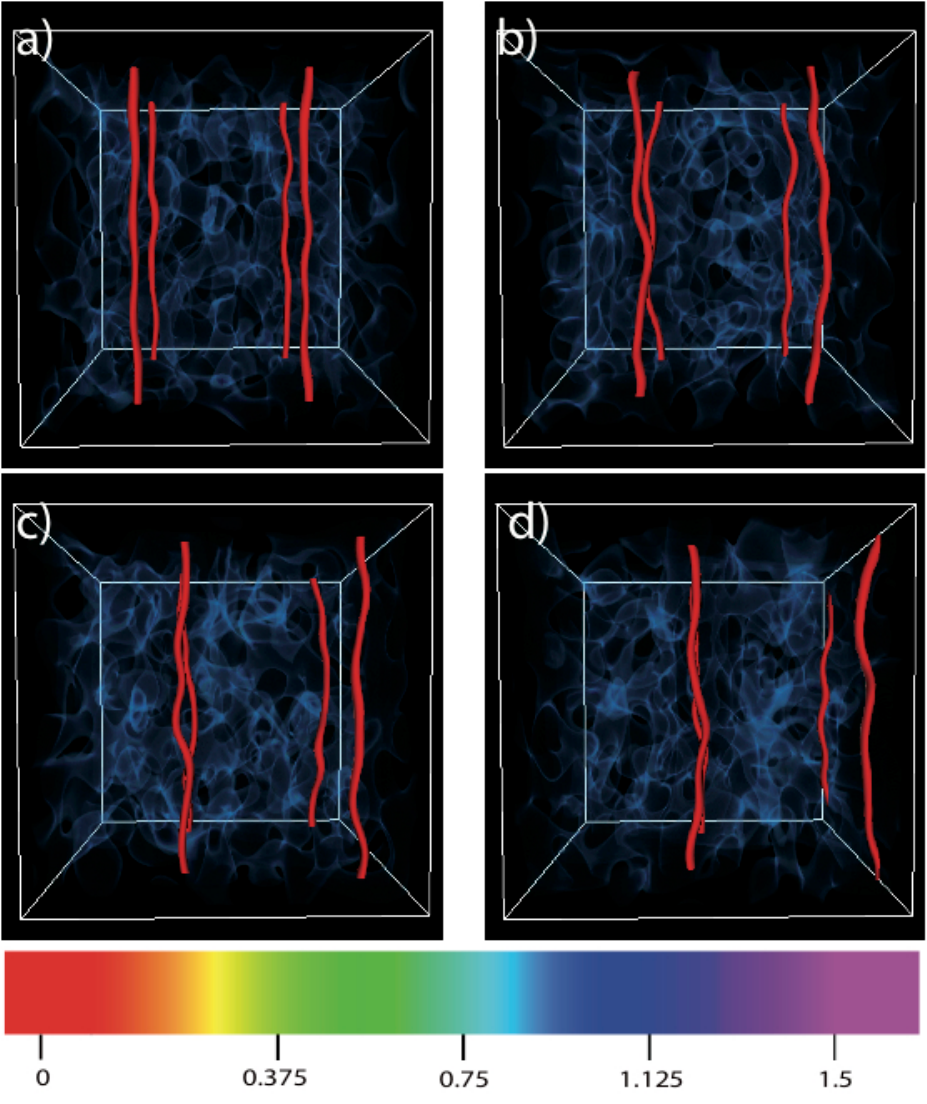}
\caption{(Color online) $3$D visualization of density at $t=0$, $40$, $60$ and $120$ at temperature $T=1$ and counterflow $W=0.4$. The grey (blue) clouds correspond to density fluctuations and the crystal-like vortex lattice is displayed in grey (red) isosurfaces}
\label{Fig:Lattice}
\end{center}
\end{figure}

\begin{figure}[htbp]
\begin{center}
\includegraphics[height=8cm]{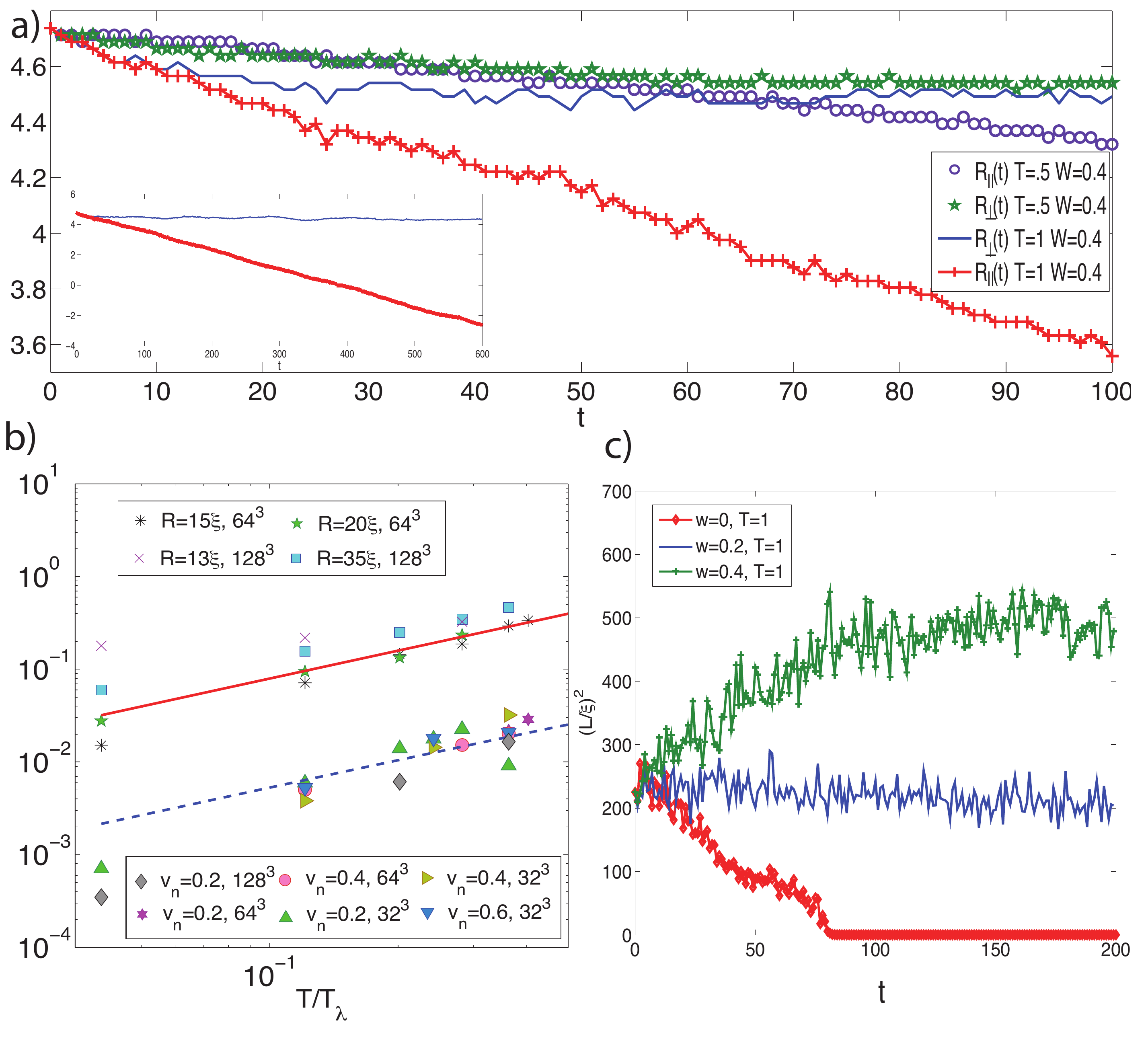}
\caption{(Color online) a) Trajectory of a straight vortex in the crystal pattern for $T=1$, $T=0.5$ and $w_{z}=.4$, at resolution $64^3$. Inset: run with $T=1$ until $t=600$.  b) Temperature dependence ($T_\lambda=3.31$) of the advection velocity $v_{\parallel}/w_{z}$ for the lattice and $\Delta v_{L}/u_{\rm i}$ for the vortex rings (resolutions $32^3$-$128^3$). The dashed line corresponds to to Eq.\eqref{Eq:alphaprime} with  $B'=0.83$ and the solid line to the theoretical prediction \eqref{Eq:AnEffect}.  c) Temporal evolution of the square of the length of the vortex ring for different values of counterflow, $T=1$ and initial radius $R=15\xi$ at resolution $64^3$.}
\label{Fig:CounterFlow}
\end{center}
\end{figure}

To numerically study the effect of counterflow on vortices we prepare an initial condition $\psi_{\rm lattice}$ consisting in a periodical array (of alternate sign) straight vortices. This initial condition is the $3$D extension of that used in ref.\cite{Nore:1994p405} to study the scattering of first sound in $2$D. The lattice is obtained using a Newton method \cite{Huepe:2000pp126-140,Tuckerman:2004Newton,Pham:2005Boundary}. It is an exact stationary solution of the (periodic) GPE. 
As the vortices are separated by a fixed distance $d=\pi$, they can be considered isolated in the limit $\xi\ll d$. Let us remark that this limit is automatically obtained  when the resolution is increased at constant $\xi k_{\rm max}$.
To include temperature effect we prepare absolute equilibria $\psi_{\rm eq}$ using SGLE with the counterflow aligned with an axis perpendicular to the vortices in $\psi_{\rm lattice}$. The initial condition $\psi=\psi_{\rm lattice}\times\psi_{\rm eq}$ is then evolved, using the TGPE. The counterflow induces a motion of the lattice as is apparent on the $3$D visualizations of the the time evolution of the density that are displayed on Fig.\ref{Fig:Lattice} (Figure obtained in the same way as Fig.\ref{Fig:meta}.b).
Several runs were performed at different resolutions (with $\xi\kmax=1.48$), temperature and counterflow values (see legend on Fig.\ref{Fig:CounterFlow}.b).

Figure \ref{Fig:CounterFlow}.a displays the temporal evolution of $(R_{\parallel},R_{\perp})$ the respectively parallel and perpendicular component of the vortex filament to the counterflow for $T=0.5$, $1$ and $w_{z}=0.4$.
The trajectories are obtained by first averaging along the direction of the vortices, then the (averaged) coordinate of the vortices is found by seeking the zero of the reduced $2d$ wavefunction. Observe that the vortex, originally located at $(\frac{3\pi}{2},\frac{3\pi}{2})$, moves in the direction of the counterflow and its velocity clearly depends on the temperature. It is apparent that a perpendicular motion is also induced at short times. This motion has two phases, the first one is related to an adaptation and makes the crystal-like lattice slightly imperfect. Then the perpendicular motion almost stops (a very small slope can be observed for long time integration). The initial phase where the parallel and perpendicular motions have similar velocities lasts longer when $\xi/d$ is decreased by increasing the resolution (data not shown). Observe that the imperfection of the lattice in the final configurations is almost equal for the two temperatures presented in Fig.\ref{Fig:CounterFlow}.a, but the parallel velocities are considerably different. Thus, the self-induced parallel velocity caused by the slight imperfection of the lattice is very small and is not driving the longitudinal motion.

We now concentrate on the measurement of $R_{\parallel}$ for which the present configuration is best suited.

$R_{\parallel}$ has a linear behavior, that allows to directly measure the parallel velocity $v_{\parallel}$. The temperature dependence of $v_{\parallel}/w_{z}$ is presented on  Fig.\ref{Fig:CounterFlow}.b  for different values of $w_z$ and $d/\xi$ (corresponding to the different resolutions).

For superfluid vortices the standard phenomenological dynamic equation of the vortex line velocity $v_{L}$ is \cite{Donne} 
\begin{equation} 
{\bf v}_{\rm L}={\bf v}_{\rm sl}+\alpha {\bf s'}\times({\bf v}_{\rm n}-{\bf v}_{\rm sl})-\alpha'{\bf s'}\times[{\bf s}'\times({\bf v}_{\rm n}-{\bf v}_{\rm sl})],\label{Eq:VortexDyn}
\end{equation}
where $s'$ is the tangent to the vortex line, ${\bf v}_{\rm sl}$ is the 
local superfluid velocity (the sum of the ambient superfluid velocity $v_{\rm s}$ and the self-induced vortex velocity $u_{\rm i}$) and $v_{\rm n}=w+v_{\rm s}$ is the normal velocity. The constants $\alpha, \alpha'$ depend on the temperature. 
Let us remark that the existence of the transverse force (related to the third term of r.h.s. in Eq.\ref{Eq:VortexDyn}) has been subject of a large debate in the low-temperature community in the last part of the 90's \cite{Thouless:1996p3049,Volovik:1996p3055,Wexler:1997p3057,Hall:1998p3040,Sonin:1998p3058,Wexler:1998p3041,Fuchs:1998p3051} and this controversy is still not resolved. 
Applied to the present case, Eq.\eqref{Eq:VortexDyn} predicts  $v_{\perp}=-\alpha w_{z}$ and $v_{\parallel}=\alpha' w_{z}$. 
The value of the constant $\alpha'$, related to the transverse force, depends on the normal density and the phonon-vortex scattering section. It can be expressed as 
\begin{equation}
\alpha'=B'\frac{\rho_{n}}{2\rho}\label{Eq:alphaprime}
\end{equation}
where $B'$ is an order one constant \cite{Donne}. A fit to the measured values of $v_{\parallel}/w_{z}$ yields $B'=0.8334$, see Fig.\ref{Fig:CounterFlow}b. We thus conclude that finite-temperature TGPE counterflow effects measured on $R_{\parallel}$ for the crystal pattern are in quantitative agreement with standard phenomenology (Eq.\eqref{Eq:VortexDyn}). 
We have seen above that the effect on $R_{\perp}$ is of the same order of magnitude that the one on $R_{\parallel}$ but only in the initial phase as long as crystal imperfection does not come into play.


\subsubsection{Vortex rings}

We now turn to study the effect of counterflow on vortex rings. The initial condition is prepared as in the previous section but with the lattice $\psi_{\rm lattice}$ replaced by a vortex ring $\psi_{\rm ring}$, that is an exact stationary (in a co-moving frame) solution of GPE. The plane containing the vortex rings of radius $R$ is perpendicular to the counterflow and the rings are numerically obtained by a Newton method \cite{Huepe:2000pp126-140,Tuckerman:2004Newton,Pham:2005Boundary}.

In the case of vortex rings the general formula \eqref{Eq:VortexDyn} yields
\begin{eqnarray}
\dot{R}&=&-\alpha(u_{\rm i}-w_z)\\
v_{\rm L}&=&v_{\rm s}+(1-\alpha')u_{\rm i}+\alpha'w_z,
\end{eqnarray}
where $u_{\rm i}$ denotes the ring velocity at zero temperature, explicitly given by $V_{\rm ring}$ in formula \eqref{Eq:vring} (replacing $R^*$ by the corresponding radius).
In the special simple case $w_z=0$, a finite-temperature contraction of the vortex ring is predicted.
This transverse effect effect was first obtained and measured by Berloff and Youd, using a finite-difference scheme version of the TGPE that exactly conserves the energy and particle number \cite{Berloff:2007p423}. 

The temporal evolution of the square of the vortex length of a ring of initial radius $R=15\xi$ at temperature $T=1$ and counterflow $w_{z}=0$, $0.2$ and $0.4$ is displayed on Fig.\ref{Fig:CounterFlow}.c. For $w=0$, the dynamics under TGPE evolution reproduces the Berloff ring contraction \cite{Berloff:2007p423}. 
The temperature dependence of the contraction obtained for $w=0$ (data not shown) quantitatively agrees with Berloff and Youd's results. 
A dilatation of vortex rings is apparent on Fig.\ref{Fig:CounterFlow}.c for $w$ larger than the measured vortex ring velocity $v_L=0.23$.  

However, $v_L$ has a very strong dependence on temperature that is also present for $w=0$. The temperature dependence of $\Delta v_{L}/u_{\rm i}$ where $\Delta v_{L}=u_{\rm i}-v_{L}$ and is displayed on Fig.\ref{Fig:CounterFlow}.b. We have checked that the velocity $v_{L}$ directly measured at $T=0$ is indeed given by $u_{\rm i}$.
Equation \eqref{Eq:VortexDyn} predicts (in the absence of counterflow) a translational velocity for the vortex ring $v_{L}=(1-\alpha')u_{\rm i}$. Observe that $\Delta v_{L}/u_{\rm i}$ is one order of magnitude above the transverse mutual friction coefficient measured on the crystal-like lattice.  
Note the presence of a large spread of the low temperature data for the ring (see the leftmost datapoints on Fig.\ref{Fig:CounterFlow}.b corresponding to $T/T_\lambda=0.04$). At very low temperatures the effect is very weak.  Thus, the corresponding measured values of $\Delta v_{L}/u_{\rm i}$ are influenced by errors on the measurement of position and velocity of the vortices that are caused by the finite-size of the mesh. In the future, these low temperature uncertainties on the determination of $\Delta v_{L}/u_{\rm i}$ could be reduced by performing runs at higher resolution together with more accurate (sub-grid) measurement of the vortex position.

\subsubsection{Anomalous translational velocity and Kelvin waves}

\begin{figure}[htbp]
\begin{center}
\includegraphics[height=9.5cm]{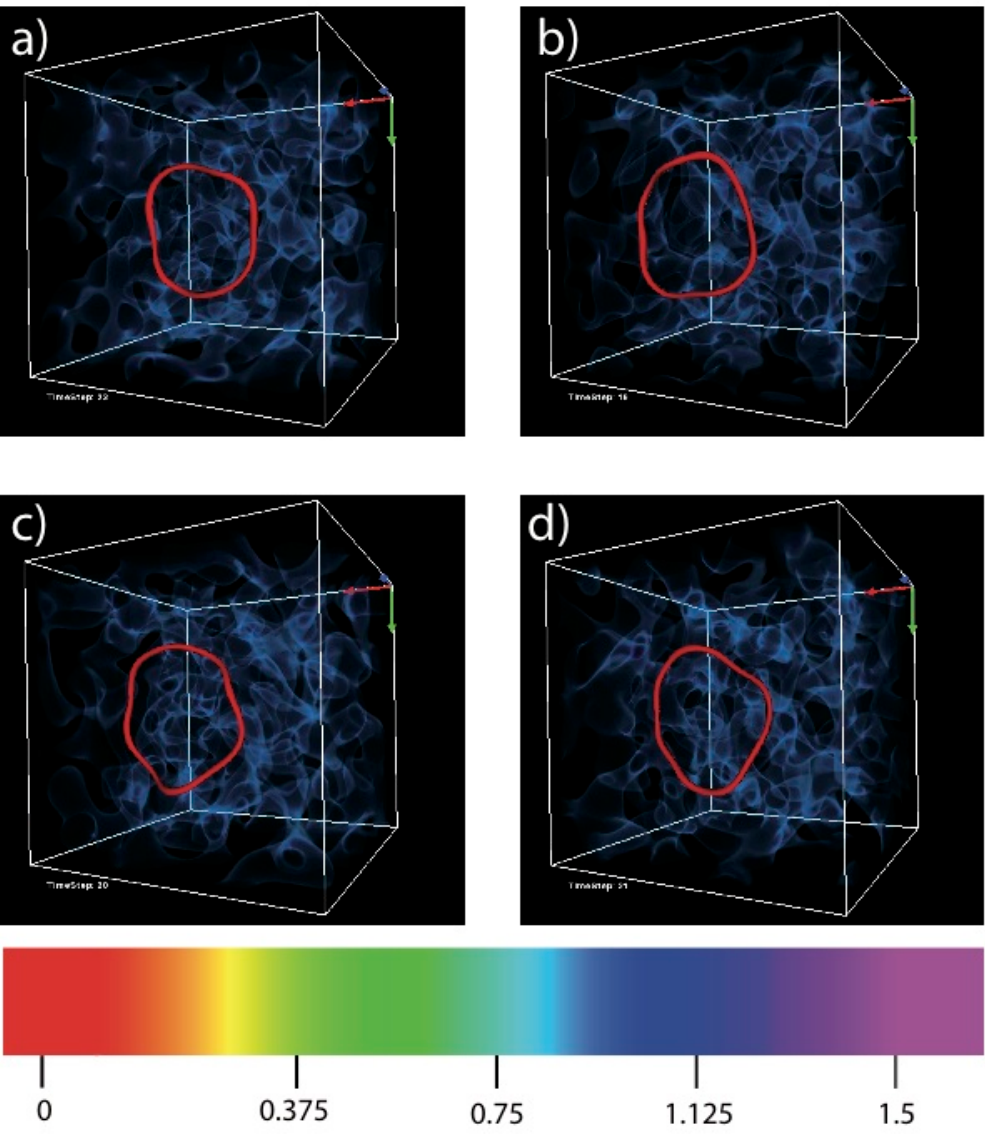}
\caption{(Color online) $3$D visualization of density at $t=18$, $19$, $20$ and $21$ at temperature $T=1$. The grey (blue) clouds correspond to density fluctuations and a vortex ring of radius $R=20\xi$ with thermally excited Kelvin waves is displayed in grey (red) isosurfaces}
\label{Fig:Kelvin}
\end{center}
\end{figure}

In this section we relate the finite temperature slowdown (see the top line of Fig.\ref{Fig:CounterFlow}.b) to the anomalous translational velocity of vortex ring with finite-amplitude Kelvin waves that was reported in refs. \cite{Kiknadze:2002p3905,Barenghi:2006p3901}.  Indeed, Kelvin waves are clearly observed in $3$D visualizations of vortex rings driven at finite-temperature by the TGPE as it is apparent on Fig.\ref{Fig:Kelvin} (obtained in the same way that Fig.\ref{Fig:meta}.b).

Following reference \cite{Barenghi:2006p3901}, Kelvin waves of amplitude $A$ and wavelength $2\pi R/N$ on a ring of radius $R$ are parametrized, in cylindrical coordinates $r$, $\phi$ and $z$, as
\begin{eqnarray}
x&=&(R+A \cos{N\phi})\cos{\phi}\label{Eq:KelvinWave1}\\
y&=&(R+A \cos{N\phi})\sin{\phi}\\
z&=&-A\sin{\phi}.\label{Eq:KelvinWave3}
\end{eqnarray}
In the limit $N\gg1$ the dispersion relation $\omega(k)$ of the Kelvin wave (\ref{Eq:KelvinWave1}-\ref{Eq:KelvinWave3}) is given by \cite{Kiknadze:2002p3905}
\begin{equation}
\omega(k)=\frac{\hbar}{2m}k^2[\ln{(\frac{8 R
}{\xi})}-a  ]\label{Eq:reldispKelvin}
\end{equation}
where $k=N/R$ and $a$ is the core model-depending constant in formula \eqref{Eq:vring}.

The anomalous translational velocity caused by an excited Kelvin wave was first reported by Kiknadze and Mamaladze \cite{Kiknadze:2002p3905} in the framework of the local induction approximation (LIA).
The effect was then obtained and numerically characterized within the Biot-Savart equation by Barenghi et al. \cite{Barenghi:2006p3901}.
The  anomalous translational velocity $v_{\rm a}$ of a vortex ring reads (in the limit $N\gg1$, see Eq.(26) of reference \cite{Kiknadze:2002p3905})
\begin{equation}
v_{\rm a}\approx u_{\rm i}(1-\frac{A^2N^2}{R^2})\label{Eq:AneffKik}
\end{equation}
where  $u_{\rm i}=V_{\rm ring}$ is the self-induced velocity \eqref{Eq:vring} without Kelvin waves.  

The variation of the energy of a vortex ring caused by a (small amplitude) Kelvin wave can be estimated as
\begin{equation}
\Delta E=\frac{dH_{\rm ring}}{dR}\frac{\Delta L}{2\pi}
\end{equation}
where $H_{\rm ring}$ is the energy given by Eq.\eqref{Eq:Hring} and the length variation $\Delta L$ produced by the Kelvin wave (\ref{Eq:KelvinWave1}-\ref{Eq:KelvinWave3}) is given, at lowest order in the amplitude $A/R$, by $\Delta L=\pi A^2 N^2/R$. Assuming equipartition of the energy of Kelvin waves with the heat bath implies $\Delta E=k_{B}T$, which yields the value of $A^2N^2/R^2$ as function of $T$:
\begin{equation}
\frac{A^2N^2}{R^2}=\frac{m^2k_BT}{\pi^2\rho_{\infty}\hbar^2R(\log{\frac{8R}{\xi}}-a)}.\label{Eq:epsk2}
\end{equation}
The equipartition law \eqref{Eq:epsk2} can also be directly obtained as the classical limit of the quantum distribution computed by Bareghi {\it et al.} \cite{Barenghi:1985p5667}, up to a redefinition of the core constant model $a$ (see Eq.(25) in reference \cite{Barenghi:1985p5667}).  
Let us remark at this point that, at low temperature and in  non-equilibrium conditions, the presence of a Kelvin wave cascade at scales between the inter-vortex distance and $\xi$ (see references \cite{Kozik:2004pKelvin,Lvov:2010}) can lead to a different dependence of the amplitude on the wavenumber.

We finally assume that the slowing down effect of each individual Kelvin wave is additive and that the waves populate all the possible modes. Kelvin waves are bending oscillations of the the quantized vortex lines, with wavenumber $k \lesssim 2 \pi / \xi$. The total number of modes can thus be estimated as
\begin{equation}
\mathcal{N}_{\rm Kelvin}\approx2\pi R/\xi .
\end{equation}
Replacing $A^2N^2/R^2$ in Eq.\eqref{Eq:AneffKik} by Eq.\eqref{Eq:epsk2} and multiplying by the total number of waves $\mathcal{N}_{\rm Kelvin}$ 
we obtain the following expression for the anomalous translational effect due to thermally exited Kelvin waves
\begin{equation}
\frac{\Delta v_{L}}{u_{\rm i}}\equiv\frac{u_{\rm i}-v_{\rm a}}{u_{\rm i}}\approx\frac{2k_{B}Tm^{2}}{\pi \rho_{{\infty}} \xi \hbar^{2}}\frac{1}{\log{\frac{8R}{\xi}}-a} \label{Eq:AnEffect}
\end{equation}

The temperature dependence of the equipartition estimate \eqref{Eq:AnEffect} of the thermal slowdown is plotted on Fig.\ref{Fig:CounterFlow}.b (top straight line). The data obtained form the measurements of the rings velocity in the TGPE runs is in very good agreement with the estimate \eqref{Eq:AnEffect}.

As discussed in refs. \cite{Davis:2001p1475,Proukakis:2008p1821} the TGPE gives a good approximation to physical (quantum) Bose-Einstein condensate (BEC) only for the modes with high phonon occupation number. In this sprit quantum effects on the Kelvin waves oscillations must also be taken into account to obtain the total slowing down effect in a BEC. 
The TGPE estimation  \eqref{Eq:AnEffect} can be adapted to weakly interacting BEC by the following considerations.

At very low temperature, because of quantum effects, only a limited range of low-wavenumber Kelvin waves are in equipartition.
This range is determined by the relation $k \le k_{\rm eq}$ with $\hbar\omega(k_{\rm eq})=k_{\rm B}T$  and the dispersion relation \eqref{Eq:reldispKelvin}, it reads:
\begin{equation}
k_{\rm eq}=\sqrt{\frac{k_{\rm B}T\,2m}{\hbar^2[\ln{(\frac{8 R}{\xi})}-a  ]}}
\end{equation}
and can also be expressed as
\begin{equation}
k_{\rm eq}=\sqrt{\frac{4\pi\, n^{2/3}}{\zeta(\frac{3}{2})^{2/3}[\ln{(\frac{8 R}{\xi})}-a  ]}}\left(\frac{T}{T_\lambda}\right)^{1/2}.
\end{equation}
where $T_\lambda$ is the Bose-Einstein condensation temperature of non-interacting particles \eqref{Eq:Tlambda} and the relation between the interatomic distance $\ell$ and the vortex-core size $\xi$ are given in Eq.\eqref{Eq:xiBEC}.

Observe that $k_{\rm eq}$ varies from $k_{\rm eq}=0$ at $T=0$ to wavenumber of order $k_{\rm eq}\sim \ell^{-1}$ at $T_\lambda$ and it is equal to $k_\xi=2\pi/\xi$ at $T^*$ defined by
\begin{equation}
T^*=8\pi^2\zeta(\frac{3}{2})^{2/3}[\ln{(\frac{8 R}{\xi})}-a  ]\left(\frac{\tilde{a}}{\ell}\right)T_\lambda.
\end{equation}
Therefore at temperatures $T¬^*<T<T_\lambda$ the energy of all Kelvin waves are in equipartition and equation \eqref{Eq:AnEffect} thus applies directly.

It is natural to suggest that an additional effect, caused by the quantum fluctuations of the amplitudes of Kelvin waves, will take place at low temperatures $T<T¬^*$ . This quantum effect can be estimated by using the standard relation for the energy of the fundamental level of a harmonic oscillator $\Delta E=\hbar\omega(k)/2$. Applied to the Kelvin waves, this relation yields the $k$-independent quantum amplitude $A_Q^2=m/4\pi^2 R\rho$.
The quantum effect can thus be estimated as the sum
\begin{equation}
\sum_{N=\mathcal{N}_{\rm Kelvin}^{eq}}^{\mathcal{N}_{\rm Kelvin}} \frac{A_Q^2 N^2}{R^2}\sim \frac{A_Q^2 {\mathcal{N}_{\rm Kelvin}}^3}{3 R^2}=\frac{2m\pi}{3\rho\xi^3}=\frac{64\pi^{5/2}}{ 3\sqrt{2}}\left(\frac{\tilde{a}}{\ell }\right)^{3/2}.
\end{equation}
The total effect is obtained superposing the thermal effect and the quantum effect and the final result is
\begin{eqnarray}
\left.\frac{\Delta v_{L}}{u_{\rm i}} \right|_{T<T^*} &=& \frac{64\pi^{5/2}}{ 3\sqrt{2}}\left(\frac{\tilde{a}}{\ell }\right)^{3/2}+\frac{(4/\sqrt{\pi} )}{\zeta(\frac{3}{2})C[\frac{R}{\xi}]^{3/2}} \left(\frac{T}{T_\lambda}\right)^{3/2}\hspace{4mm}\\
\left.\frac{\Delta v_{L}}{u_{\rm i}} \right|_{T>T^*}  &=&
   \frac{8\sqrt{2\pi}}{\zeta(\frac{3}{2})^{2/3}C[\frac{R}{\xi}] }\left(\frac{\tilde{a}}{\ell}\right)^{1/2}\frac{T}{T_\lambda}
\label{Eq:AnEffectQuant}
\end{eqnarray}
where $C[R/\xi]=\log  \left(\frac{8 R}{\xi }\right)-a$.

In the case of superfluid Helium, where $\tilde{a}\sim \ell$, the GPE description is only expected to give qualitative predictions and, at best, order of magnitude estimates (see ref.\cite{Donne}). 
It is thus difficult to extend the above considerations, obtained in the case of weakly interacting BEC with $\tilde{a}\ll\ell$,
to Helium. 

Nevertheless the results obtained above in the weakly interacting case strongly suggest the presence of new slowing down effects, not included in the usual mutual friction descriptions of Helium that predicts $\frac{\Delta v_{L}}{u_{\rm i}} \sim\rho_n/\rho\sim (T/T_\lambda)^4$. The new effects, because of their temperature dependence (see Eq.\eqref{Eq:AnEffectQuant}), should be dominant at low-temperature. 

The zero-temperature quantum slowdown is independent of the ring diameter and the finite temperature effects are stronger for small rings.
Time of flight measurements  of vortex rings in $^4{\rm He}$ could be used to determine the translational velocity. 
The effect could also be studied in ultra-cold atomic gases BEC. For these systems the effect of the inhomogeneity of the superfluid should be taken into account \cite{Krstulovic:2010p5967}.

\section{Conclusions\label{Sec:Conclusion}}

In summary, our main results were obtained by making use of a stochastically forced Ginzburg-Landau equation (SGLE) that permits to efficiently obtain and control truncated Gross-Pitaevskii absolute equilibrium.
This allowed us to show that the condensation transition observed in references \cite{Davis:2001p1475,Connaughton:2005p1744,During:2009PhysicaD} corresponds to a standard second-order transition described by the $\lambda \phi^4$ theory.

We also found that thermodynamic equilibrium can be obtained by a direct energy cascade, in a way similar to that of Cichowlas et al.\cite{Cichowlas:2005p1852}, accompanied by vortex annihilation as a prelude to final thermalization. Increasing the amount of dispersion of the system a slowdown of the energy transfer was produced inducing a partial thermalization independently of the truncation wavenumber.  This new thermalization regime opens up an avenue to a further investigation of vortex dynamic in co-flowing finite-temperature superfluid turbulence. In this context it would be interesting to study in the future, using a much higher resolution than in the present work, the dispersive bottleneck. In particular to investigate the possibility of the coexistence of a well established turbulent Kolmogorov cascade followed by a dispersive-induced partial-thermalization zone.

Using the SGLE in the presence of a counterflow we observed that the counterflow can block the contraction of vortex rings reported by Berloff and Youd \cite{Berloff:2007p423} and also induce a dilatation. We directly measured the mutual friction coefficient related to the transverse force. An unexpected result was found by immersing a vortex ring in a finite-temperature bath: a strong dependence of the translational velocity in the temperature was observed. This effect was an order of magnitude above the transverse mutual friction effect. We explained this effect by relating it to to the anomalous translational velocity due to finite amplitude Kelvin waves that was previously found by Kiknadze and Mamaladze \cite{Kiknadze:2002p3905} and Barenghi et al \cite{Barenghi:2006p3901}. Assuming equipartition of the energy of the Kelvin waves with the heat bath yields a formula that gives a very good quantitative estimate of the numerically observed effect.
This new formula also gives an experimentally-testable quantitative prediction for the thermal slowdown of vortex rings in weakly interacting Bose-Einstein condensates and superfluid $^4{\rm He}$.
In this context, it would be interesting in the future to study (using a higher resolution than in the present work) the vortex dynamic of counter-flowing finite-temperature superfluid turbulence.
Note that, in the context of BEC (where experiments are performed within a confining potential) the wavefunction $\psi$ must be expanded using another basis of orthogonal independent functions than the Fourier modes (e.g. the eigenfunctions of the harmonic oscilator), see ref.\cite{BlakieDavis05} where the apparent arbitrariness of the truncation parameter is also discussed. About this last point, also see the discussion around Eqs.\eqref{Eq:bogu} and \eqref{Eq:Tstar} at the end of Sec. \ref{SubSection:SelfTrunc} of the present work about the upper limit $k_{\rm eq}$ of the equipartition range that follows from the quantization of phonons in a physical BEC.

The TGPE dynamics was thus found to contain many physically sound phenomena of finite-temperature superflows. This strongly suggests the possibility to obtain the propagation of second sound waves in the TGPE. Some preliminary results support this conjecture (data not shown), however very high resolutions seem to be needed and this will be the subject of a future work.

\section*{Acknowledgments}

We acknowledge useful scientific discussions with G. D\"{u}ring and S. Rica. The computations were carried out at IDRIS (CNRS).
\appendix

\section{Conservation Laws and Dealiasing \label{Ap:Des&Cons}}

In the standard incompressible Euler case, for quadratic nonlinearities and quadratic invariants, the system can be correctly dealiased using the $2/3-$rule that consists in truncation for wavenumber $|{\bf k}|<k_{\rm max}=N/3$, where $N/2$ is the largest wavenumber of the discrete system.  With this procedure, one third of the available modes are not used. Such discrete dealiased pseudo-spectral system exactly conserve the quadratic invariant and is therefore identical to the original Galerkin truncated system.

In the TGPE case, the problem is more complicated because the equation is cubic and the invariants are quartic. Let us first recall Parseval's theorem that states
$\int\,d^3x f({\bf x})g^*({\bf x})=V\sum_{\bf k} \hat{f}_{\bf k}\hat{g}_{\bf k}^*$, 
 where $\hat{f}_{\bf k}$ and $\hat{g}_{\bf k}$ are the Fourier transform of $f$ and $g$. This identity remains valid in truncated systems and it holds whether the functions are dealiased or not. The integration by parts formula is a consequence of Parseval's theorem:
\begin{eqnarray*}
\int d^3x \,f\pder{g^*}{x_j}&=&V\sum_{\bf k} -ik_j\hat{f}_{\bf k}\hat{g}_{\bf k}^*=-\int d^3x\, \pder{f}{x_j} g^*.\label{Eq:Parseval2}
\end{eqnarray*}
Remark that the product rule $(fg)'=f'g+fg'$ is only valid if the fields are dealiased.

The conservation of the total number of particles is directly obtained using the GPE \eqref{Eq:GPE}
\begin{equation*}
\frac{dN}{dt}=\int \,d^3x(\dot{\psi}\bar{\psi}+\psi\dot{\bar{\psi}})=\frac{i\hbar}{2m}\int \,d^3x(\bar{\psi}\nabla^2\psi-\psi\nabla^2\bar{\psi})=0.
\end{equation*}
where the last equality is a consequence of the Parseval identity and is thus true independently of dealiasing. Similar relations lead to the conservation of the energy $H$.

Using the dealiased TGPE \eqref{Eq:TGPEphys} the conservation law for the momentum reads
\begin{equation}
\frac{dP_j}{dt}=2g\int \,d^3x\left[\left(\partial_j\mathcal{P}_{\rm G} [|\psi|^2]\right)|\psi|^2+\mathcal{P}_{\rm G} [|\psi|^2]\partial_j|\psi|^2\right].\label{Eq:desP4}
\end{equation}
If $\psi$ is dealiased the $2/3-$rule implies that
\begin{eqnarray*}
\int \,d^3x(\mathcal{P}_{\rm G} [|\psi|^2]\bar{\psi})\partial_j\psi\label{Eq:GalerkinProp1}&=&\int \,d^3x\mathcal{P}_{\rm G} \left[\mathcal{P}_{\rm G} [|\psi|^2]\bar{\psi}\right]\partial_j\psi\\
\partial_j\left(\mathcal{P}_{\rm G} [|\psi|^2]\bar{\psi}\right)&=&\left(\partial_j \mathcal{P}_{\rm G} [|\psi|^2]\right)\bar{\psi}+\mathcal{P}_{\rm G} [|\psi|^2]\partial_j\bar{\psi}\label{Eq:GalerkinProp2}\\
\partial_j|\psi|^2&=&\psi\partial_j\bar{\psi}+\partial_j\psi\bar{\psi},\label{Eq:GalerkinProp3}
\end{eqnarray*}
it follows that $\frac{dP_j}{dt}=0$. Without a Galerkin projector in Eq.\eqref{Eq:TGPEphys} the aliased field would obey $ \left(|\psi|^2\bar{\psi}\right)\partial_j\psi+\left(|\psi|^2\psi\right)\partial_j\bar{\psi}\neq\partial_j(|\psi|^4)$ and the conservation of momentum would therefore be lost.

Conservation of $N$, $H$ and ${\bf P}$ can be numerically checked by using absolute equilibria with non-zero momentum. The conservation of ${\bf P}$ is ensured only if the system is dealiased. The error of aliased runs grow up to a $50\%$ in a few units of time and is independent of the time-step (data not shown). 
We thus believe that it would important to explicitly check the conservation of momentum when using finite-difference schemes, even if  they exactly conserve the energy and the particle number.

\section{Low-temperature calculation of thermodynamic functions\label{App:Low-T}}

We are interested in computing the grand partition function $\mathcal{Z}$ in Eq.\eqref{Eq:defZformal} where $F=H-\mu N-{\bf W}\cdot{\bf P}$ is written in terms of Fourier amplitudes as
\begin{eqnarray}
\nonumber\frac{H}{V}&=&\sum_{\bf k} \frac{\hbar ^2k^2}{2 m} |A_{\bf k}|^2+\frac{ g}{2}\sum A_{{\bf k_3+k_1}}^*A_{{\bf k_2}}A_{{\bf k_4+k_2}}^*\delta_{{\bf k_3},{\bf -k_4}} \label{energie2}\\
&&\\
N &=& V \sum_{\bf k} |A_{\bf k}|^2\\
P_j&=&\sum_{\bf k}\hbar k_j |A_{\bf k}|^2V\label{Eq:PinFourier}
\end{eqnarray}
where  $A_{\bf k}=0$ if $k\ge k_{\rm max}$ and the second sum in $H$ is over ${\bf k_1}$, ${\bf k_2}$, ${\bf k_3}$, ${\bf k_4}$.

The saddle-point is determined by the condition $\pder{F}{A_{\bf k}^*}-\mu_{0}A_{\bf 0}V\delta_{\bf k,0}=0$ which, separately written for ${\bf k}={\bf 0}$ and ${\bf k}\neq{\bf 0}$, explicitly reads
\begin{eqnarray}
(g| A_{\bf 0}|^2-\mu+\mu_{0})A_{\bf 0}+2g\sum_{{\bf k_1\neq0}}A_{\bf 0}|A_{\bf k_1}|^2\hspace{1.2cm}&&\label{Eq:col2}\\
\nonumber+g\sum_{{\bf k_1},{\bf k_2}\neq{\bf 0}}A_{\bf k_1}A^*_{\bf k_2-k_1}A_{\bf - k_2}\hspace{0cm}&=&0\\
\frac{\hbar^2k^2}{2m} A_{\bf k}-\mu A_{\bf k}-\hbar{\bf W}\cdot{\bf k}\,A_{\bf k}\hspace{3cm}&&\label{Eq:col3}\\
\nonumber+\bet\sum_{{\bf k_1},{\bf k_2}\neq0}A_{\bf k_1}A^*_{\bf k_2+k_1}A_{\bf k+ k_2}\hspace{0cm}&=&0
\end{eqnarray}
from which Eq.\eqref{Eq:solcol} follows.

To diagonalize $F=H-\mu N-{\bf W}\cdot{\bf P}$ we first apply the Bogoliubov transformation to $H-\mu N$ and then show that ${P}$ is also diagonal in this basis. Replacing $B_{\bf p}$, defined by the transformation \eqref{Eq:transBogo}, in $H-\mu N$ (recall that ${\bf p}=\hbar{\bf k}$) and then imposing the diagonalization determines the coefficient $L_{p}$:
\begin{equation}
L_{ p}=\frac{-2  |A_{\bf 0}|^2 g-\frac{p^2}{2
   m}+\mu+\epsilon(p)}{ |A_{\bf 0}|^2 g}\label{Eq:eqLp}
\end{equation}
where $\epsilon(p)$ is given by
\begin{equation}
\epsilon(p)=\sqrt{\left(2  |A_{\bf 0}|^2 g+\frac{p^2}{2 m}-\mu \right)^2- |A_{\bf 0}|^4 g^2}.
\end{equation}
The dispersion relation \eqref{Eq:RelDispSaddle} is obtained by replacing $|A_{\bf 0}|^2$ by its saddle-point value Eq.\eqref{Eq:solcol}. 

We now express ${\bf P}$ in the Bogoliubov base. Using (\ref{Eq:transBogo}) directly yields
\begin{equation}
|A_{\bf p}|^2=|u_{ p}|^2|B_{\bf p}|^2+|v_{ p}|^2|B_{\bf -p}|^2+({u^*}_{ p}{v^*}_{ p}B_{\bf p}B_{\bf -p}+c.c).\label{Eq:As}
\end{equation}
Replacing Eq.\eqref{Eq:As} in the definition of ${\bf P}$ \eqref{Eq:PinFourier}, the last two terms vanish by symmetry and using the relation $|u_{ p}|^2-|v_{ p}|^2=1$, the momentum \eqref{Eq:PinFourier} reads
${\bf P}=\sum_{\bf p} {\bf p} |B_{\bf p}|^2V$. 
Formula \eqref{Eq:Fbogu} is then finally obtained by gathering $H-\mu N$ and ${\bf W}\cdot{\bf P}$.

The mean value of the condensate amplitude is obtained as $V\overline{|A_{\bf 0}|^2}=-\left.\frac{\partial\Omega}{\partial\mu_{0}}\right|_{\mu_{0}=0}$. All the thermodynamic variables are directly generated by first putting $\mu_{0}=0$ in \eqref{Eq:PartialOmegalowT} and then by differentiation, using relation \eqref{Eq:dPartialOmega}. The fluctuations of the number of particles are computed as $ \overline{\delta N^2}=-\beta^{-1}\frac{\partial^2 \Omega}{\partial\mu^2}$. These quantities are explicitly listed below.
\begin{eqnarray}
\nonumber\overline{|A_{\bf 0}|^2}&=&\frac{\mu}{g}-\frac{\mathcal{N}}{V\beta \mu}f_{0}\left[\frac{4 m \mu
   }{P_{\rm max}^2}\right]\\
\nonumber   \bar{ p} &=& \frac{\mu ^2}{2 g}+\frac{\mathcal{N}}{ V \beta }\left(\frac{2}{3}- f\left[\frac{4 m \mu
   }{P_{\rm max}^2}\right]+\frac{2}{3}\frac{2 w^2 m^2}{P_{\rm max}^2} f'\left[\frac{4 m \mu
   }{P_{\rm max}^2}\right]\right)\label{Eq:PreslowT}\\
\nonumber\bar{ N}&=&\frac{V \mu }{g}-\frac{\mathcal{N}}{\beta}\left(\frac{3}{2\mu} f\left[\frac{4 m \mu
   }{P_{\rm max}^2}\right]- \frac{8w^2 m^3}{P_{\rm max}^4}f_2\left[\frac{4 m \mu
   }{P_{\rm max}^2}\right]\right) \label{Eq:NtotlowT}\\
\nonumber  S &=&\mathcal{N}\left(f\left[\frac{4 m \mu
   }{P_{\rm max}^2}\right](1+ \frac{2 w^2 m}{4 \mu})-\log{\left[\frac{\beta\epsilon(P_{\rm max};\mu)}{e^{-\frac{5}{3}}}\right]}\right) \\
\nonumber   \lambda_{\mathcal{N}}&=& \beta^{-1}\log{[\beta\epsilon(P_{\rm max};\mu)]}-\frac{1}{3\beta}\frac{2 w^2 m^2}{P_{\rm max}^2}\frac{1}{1+\frac{4 m \mu
   }{P_{\rm max}^2}}\\
\nonumber  \bar{ P_z}&=& \frac{ \mathcal{N}}{\beta}   \frac{ w m}{\mu}  f\left[\frac{4 m \mu
   }{P_{\rm max}^2}\right] + \frac{3 \mathcal{N}}{10\beta}   \frac{ w^3 m^2}{ \mu^2} f_1\left[\frac{4 m \mu
   }{P_{\rm max}^2}\right]\\
\nonumber  \overline{\delta N^2}&=&\frac{V}{g\beta}+\frac{3\mathcal{N}}{4\beta^2\mu^2}f_1\left[\frac{4 m \mu
   }{P_{\rm max}^2}\right],\\
&&   \label{Eq:LowTFunctions}\\
f[z]&=&z-z^{3/2} \cot ^{-1}\left(\sqrt{z}\right)\label{Eq:f}\\
f_{0}[z]&=&3(z+3f[z])/4\label{Eq:f0}\\
f_1[z]&=&\frac{z}{z+1}-f(z)\label{Eq:f1}\\
f_2[z]&=&\frac{d}{dz}(f[z]/z)\label{Eq:f2}
\end{eqnarray}

The dependence of the entropy on the phase-space normalization constant is manifested by the presence of the logarithm term in $S$ and $\lambda_{\mathcal{N}}$. Notice that the function $S+\beta\lambda_{\mathcal{N}}$ is, however, completely defined. Also note that the pressure $p$ must be computed, by definition, at  constant total number of modes $\mathcal{N}$. All the thermodynamic relations discussed in section \ref{SubSec:Thermo} can be explicitly checked on the low-temperature expressions.  The previous formulae are represented as the solid lines that are confronted with the SGLE numerically generated data in Fig.\ref{Fig:Scan}.a-b.


\begin{thebibliography}{10}

\bibitem{Landau6Course}
L.~D. Landau and L.~M. Lifshitz,
\newblock {\em Course of Theoretical Physics, Volume VI: Fluid Mechanics}.
\newblock Butterworth-Heinemann, 2 edition, January 1987.

\bibitem{Vinenxxx}
W.~F. Vinen,
\newblock { Proc. R. Soc. Lond. A}  \textbf{242}, 493 (1957).

\bibitem{Proukakis:2008p1821}
N.P. Proukakis and B. Jackson,
\newblock { J. Phys. B: At. Mol. Opt. Phys.} \textbf{41}, 203002 (2008 ). 

\bibitem{Donne}
R.~J. Donnelly,
\newblock {\em Quantized Vortices in Helium II}.
\newblock Cambridge Univ. Press, 1991.

\bibitem{Nore:1997p1333}
C.~Nore, M.~Abid and M.E.~Brachet,
\newblock { Phys. Rev. Lett.} \textbf{78}, 3896 (1997).

\bibitem{Nore:1997p1331}
C.~Nore, M.~Abid and M.E.~Brachet,
\newblock {Phys. Fluids} \textbf{9}, 2644 (1997).

\bibitem{Kobayashi:2005p4037}
M.~Kobayashi and M.~Tsubota,
\newblock { Phys. Rev. Lett.} \textbf{94}, 065302 (2005).

\bibitem{Yepez:2009p5107}
J.Yepez, G. Vahala, L. Vahala and M. Soe,
\newblock { Phys. Rev. Lett.} \textbf{103}, 084501 (2009).

\bibitem{Abid:1998p3893}
M.~Abid, M.E.~Brachet, J.~Maurer, C.~Nore and P.~Tabeling,
\newblock { Eur. J. Mech. B-Fluid.} \textbf{17}, 665 (1998).

\bibitem{Maurer:1998p4365}
J.~Maurer and P.~Tabeling,
\newblock {Europhys. Lett.} \textbf{43}, 29 (1998).

\bibitem{Bewley:xxx}
G.~P. Bewley, M.~S. Paoletti, K.~R. Sreenivasan and D.~P. Lathrop,
\newblock {PNAS} \textbf{105}, 13707 (2008).

\bibitem{Paoletti:1545}
M.~S. Paoletti, M.~E. Fisher, K.~R. Sreenivasan and D.~P. Lathrop,
\newblock {Phys. Rev. Lett.} \textbf{101}, 154501 (2008).

\bibitem{Davis:2001p1475}
M.J.~Davis, S.A.~Morgan and K.~Burnett,
\newblock {Phys. Rev. Lett.} \textbf{87}, 160402 (2001).

\bibitem{ZNG99}
E.~Zaremba, T.~Nikuni and A.~Griffin, 
\newblock {J. Low Temp. Phys.} \textbf{116}, 277 (1999).

\bibitem{JPB07}
B.~Jackson, N. P.~Proukakis and C. F.~Barenghi,
\newblock {Phys. Rev. A} \textbf{75}, 051601, (2007).

\bibitem{JPBZ09}
B.~Jackson,, N. P.~Proukakis, C. F.~Barenghi, and E.~Zaremba,
\newblock {Phys. Rev. A} \textbf{79}, 053615 (2009). 

\bibitem{LEE:1952p4100}
T.D. Lee,
\newblock {Quart. Appl. Math.}, 10(1):69 (1952).

\bibitem{KRAICHNAN:1955p3039}
R.H. ~Kraichnan,
\newblock {J. Acoust. Soc. Am.} \textbf{27}, 438 (1955).

\bibitem{KRAICHNARH:1973p2909}
R.H. ~Kraichnan,
\newblock { J. Fluid Mech.} \textbf{59}, 745 (1973).

\bibitem{OrszagHouches}
S.A. Orszag,
\newblock {\em Statistical Theory of Turbulence}.
\newblock in, Les Houches 1973: Fluid dynamics, R.\ Balian and J.L. Peube eds.
  Gordon and Breach, New York, 1977.

\bibitem{Cichowlas:2005p1852}
C.~Cichowlas, P.~Bona\"iti, F.~Debbasch and M.~Brachet,
\newblock {Phys. Rev. Lett.} \textbf{95}, 264502 (2005).

\bibitem{Bos:2006p62}
W.~J.~T Bos and J.P. Bertoglio,
\newblock { Phys. Fluids.} \textbf{18}, 071701 (2006).

\bibitem{Krstulovic:2008p428}
G.~Krstulovic and M.~Brachet,
\newblock {Physica D} \textbf{237}, 2015 (2008).

\bibitem{Krstulovic:2009p1876}
G.~Krstulovic, P.~D. Mininni, M.~E. Brachet and A.~Pouquet,
\newblock {Phys. Rev. E} \textbf{79}, 056304 (2009).

\bibitem{Frisch:2008p1877}
U. Frisch, S. Kurien, R. Pandit, W. Pauls, S.~S. Ray,
  A. Wirth and J.Z. Zhu,
\newblock {\em Phys. Rev. Lett.} \textbf{101}, 144501 (2008).

\bibitem{Krstulovic:2009IJBC}
G.~Krstulovic, C.~Cartes, M.~Brachet, and E.~Tirapegui,
\newblock { IJBC} \textbf{19}, 3445 (2009).

\bibitem{Connaughton:2005p1744}
C~Connaughton, C~Josserand, A~Picozzi, Y~Pomeau and S~Rica,
\newblock {Phys. Rev. Lett.} \textbf{95}, 263901 (2005).

\bibitem{During:2009PhysicaD}
A.~Picozzi G.~D{\"u}ring and S. Rica,
\newblock { Physica D} \textbf{238}, 1524 (2009).

\bibitem{Berloff:2002p3383}
N.G.~Berloff and B.V.~Svistunov,
\newblock { Phys. Rev. A} \textbf{66}, 013603 (2002).

\bibitem{Berloff:2007p423}
N.~G. Berloff and A.~J. Youd,
\newblock {Phys. Rev. Lett.}  \textbf{99},145301 (2007).

\bibitem{ZinnJustin:2007p3931}
J. Zinn-Justin,
\newblock {\em Phase Transitions and Renormalisation Group}.
\newblock Oxford University Press, USA, August 2007.

\bibitem{JAmit:1978p4003}
D.~J. Amit and V. Martin-Mayor,
\newblock {\em Field Theory; The Renormalization Group and Critical Phenomena}.
\newblock {World Scientific Publishing Company}, June 2005.

\bibitem{Abid:2003p3895}
M.~Abid, C.~Huepe, S.~Metens, C.~Nore, C.T.~Pham, L.S.~Tuckerman and M.~Brachet,
\newblock { Fluid. Dyn. Res.} \textbf{33}, 509 (2003).

\bibitem{Got-Ors}
D.~Gottlieb and S.~A. Orszag,
\newblock SIAM, Philadelphia, 1977.

\bibitem{Landau5Course}
L.~D. Landau and L.~M. Lifshitz,
\newblock {\em Course of Theoretical Physics, Volume V: Statistical Physics
  (Part 1)}.
\newblock Butterworth-Heinemann, August 1996.

\bibitem{Van-Kampen-Chem}
N.~G.~van. Kampen,
\newblock {\em Stochastic processes in physics and chemistry}.
\newblock North-Holland ; sole distributors for the USA and Canada, Elsevier
  North-Holland, Amsterdam ; New York : New York (1981).

\bibitem{LivreVertET}
F. Langouche, D. Roekaerts, and E. Tirapegui,
\newblock {\em Functional integration and semiclassical expansions}.
\newblock D Reidel Pub Co, Jan 1982.

\bibitem{Mathews:1970p5033}
J. Mathews and R.~Lee Walker,
\newblock {\em Mathematical methods of physics‎}.
\newblock W. A. Benjamin, 1970.

\bibitem{Landau9Course}
E.~M. Lifshitz and L.~P. {P}itaevskii,
\newblock {\em Course of Theoretical Physics, Volume IX: Statistical Physics
  (Part 2)}.
\newblock Butterworth-Heinemann, January 1980.

\bibitem{Lipa:2003p2679}
J.A.~Lipa, J.A.~Nissen, D.A.~Stricker,D.R.~Swanson and T.C.P.~Chui,
\newblock {Phys. Rev. B} \textbf{68}, 174518 (2003).

\bibitem{krstulovic-2009}
G. Krstulovic and M. Brachet,
\newblock { Phys. Rev. Lett.} \textbf{105}, 129401 (2010).

\bibitem{Fermi:1955p3871}
E.~Fermi, J.~Pasta and S.~Ulam,
\newblock { LASL Report LA-1940} (1955).

\bibitem{Henn:2009p5720}
E.~A.~L Henn, J.~A Seman, G~Roati, K.~M.~F Magalhaes,  and V.~S Bagnato,
\newblock { Phys. Rev. Lett.} \textbf{103}, 045301 (2009).

\bibitem{Krstulovic:2010p5966}
G. Krstulovic and M. Brachet,
\newblock arXiv:1007.4441v2 [physics.flu-dyn], Jul 2010.

\bibitem{Huepe:1999p1467}
C.~Huepe, S.~Metens, G.~Dewel, P.~Borckmans and M.E.~Brachet,
\newblock {Phys. Rev. Lett.}, \textbf{82}, 1616 (1999).

\bibitem{Gardiner1996Handbook}
C.~W. Gardiner,
\newblock {\em Handbook of Stochastic Methods: For Physics, Chemistry and the
  Natural Sciences (Springer Series in Synergetics)}.
\newblock Springer, November 1996.

\bibitem{Winiecki:1999p607}
T.~Winiecki, J.F.~McCann and C.S.~Adams,
\newblock {Europhys. Lett.} \textbf{48}, 475 (1999).

\bibitem{Huepe:2000pp126-140}
C.~Huepe and M.E.~Brachet,
\newblock { {Physica D}} \textbf{140},126 (2000).

\bibitem{Tuckerman:2004Newton}
L.S. Tuckerman, C. Huepe and M.E. Brachet,
\newblock {Nonlin. Phen. and Cmplx. Syst.} \textbf{9}, 75 (2004).

\bibitem{Pham:2005Boundary}
C.T.~Pham, C.~Nore and M.E.~Brachet,
\newblock {{Physica D}} \textbf{210}, 203 (2005).


\bibitem{Nore:1994p405}
C.~Nore, M.E.~Brachet, E.~Cerda and E.~Tirapegui,
\newblock {Phys. Rev. Lett.} \textbf{72}, 2593 (1994).

\bibitem{Thouless:1996p3049}
D.J.~Thouless, P.~Ao and Q.~Niu,
\newblock { Phys. Rev. Lett.} \textbf{76}, 3758 (1996).

\bibitem{Volovik:1996p3055}
G.E.~Volovik,
\newblock { Phys. Rev. Lett.}  \textbf{77}, 4687 (1996).

\bibitem{Wexler:1997p3057}
C.~Wexler,
\newblock { Phys. Rev. Lett.} \textbf{79}, 1321 (1997).

\bibitem{Hall:1998p3040}
H.E.~Hall and J.R.~Hook,
\newblock { Phys. Rev. Lett.} \textbf{80}, 4356 (1998).

\bibitem{Sonin:1998p3058}
E.B.~Sonin,
\newblock { Phys. Rev. Lett.} \textbf{81}, 4276 (1998).

\bibitem{Wexler:1998p3041}
C.~Wexler, D.J.~Thouless, P.~Ao and Q.~Niu,
\newblock { Phys. Rev. Lett.} \textbf{80}, 4357 (1998).

\bibitem{Fuchs:1998p3051}
J.~Fuchs, G.~Malka, J.~C. Adam, E.~Amiranoff, S.~D. Baton, N.~Blanchot, A.~H{\'e}ron,
  G.~Laval, J.~L. Miquel, P.~Mora, H.~P{\'e}pin and C.~Rousseaux,
\newblock { Phys. Rev. Lett.} \textbf{81}, 4275 (1998).

\bibitem{Kiknadze:2002p3905}
L.~Kiknadze and Y.~Mamaladze,
\newblock { J. Low Temp. Phys.} \textbf{126}, 321 (2002).

\bibitem{Barenghi:2006p3901}
C.~F. Barenghi, R.~Hanninen and M.~Tsubota,
\newblock { Phys. Rev. E} \textbf{74}, 046303 (2006).

\bibitem{Barenghi:1985p5667}
C.F.~Barenghi, R.J.~Donnelly and W.F.~Vinen,
\newblock { Phys. Fluids.} \textbf{28}, 498 (1985).

\bibitem{Kozik:2004pKelvin}
E. Kozik and B. Svistunov,
\newblock { Phys. Rev. Lett.} \textbf{92}, 035301 (2004).

\bibitem{Lvov:2010}
V.S. L'vov and S. Nazarenko,
\newblock{\em JETP Lett.} \textbf{9}, 8 (2010)

\bibitem{Krstulovic:2010p5967}
G. Krstulovic and M. Brachet,
\newblock	 arXiv:1006.4315v2 [cond-mat.stat-mech], (2010).

\bibitem{BlakieDavis05}
P.B.~ Blakie and M.J.~ Davis,  
\newblock { Phys. Rev. A} \textbf{72}, 063608 (2005).

\bibitem{Zagrebnov:2001p3303}
V.A.~Zagrebnov and J.B.~Bru,
\newblock { Phys. Rep.} \textbf{350}, 292 (2001).

\end{thebibliography}


\end{document}